\documentclass[superscriptaddress,nobibnotes,nofootinbib,amsmath,amssymb,notitlepage,prb,aps,longbibliography,10pt]{revtex4-2}

\usepackage{bm,braket}
\usepackage[toc,page]{appendix}
\usepackage{comment}
\usepackage{dcolumn}
\usepackage{amsfonts}
\usepackage{amsmath}
\usepackage{amssymb}
\usepackage{graphicx,color,hyperref}
\graphicspath{{./img/}}
\usepackage[caption=false]{subfig}
\usepackage{url}
\hypersetup{colorlinks=true, linkcolor=blue, citecolor=blue, urlcolor=blue} 
\usepackage{xcolor}
\usepackage{bbm}
\usepackage{physics}
\usepackage{dcolumn}
\usepackage{enumitem}
\usepackage{array, multirow}
\usepackage[T1]{fontenc}

\newcommand\swap[0]{{\;\mathrm{SWAP}}}
\DeclareMathAlphabet{\mathcal}{OMS}{cmsy}{m}{n}

\makeatletter
\renewcommand*{\fnum@figure}{{\normalfont\bfseries \figurename~\thefigure}}
\makeatother

\allowdisplaybreaks

\graphicspath{{./img/}}

\begin{document}

\title{Symmetry Resolved Multipartite Entanglement Entropy}

\newcommand{\UoM}{Department of Physics and Astronomy, University of Manchester, Oxford Road, Manchester M13 9PL, UK}
\author{Ashwat Jain}
\email{ashwat.jain@postgrad.manchester.ac.uk}
\affiliation{\UoM}

\date{\today}
\begin{abstract}
We perform the symmetry resolution of a multipartite entanglement measure, namely the global entanglement $Q$ introduced by Meyer and Wallach [2002, J. of Math. Phys., 43, pp. 4273] for all systems of distinguishable particles hosting a locally acting symmetry. For an ensemble of Haar random states we find agreement with equipartition, with leading order  behaviour and finite size corrections which follow a power law scaling with the number of local degrees of freedom. Implications of this result for the general symmetry-resolved multipartite entanglement paradigm are discussed and some possible experimental verification methods are presented.
\end{abstract}
\maketitle

\section{Introduction}
\label{sec:intro}

Entanglement~\cite{einstein1935,schrodinger1935}, is a fundamental and practical resource leveraged across a large number of fields~\cite{horodecki2009,amico2008}. For example, in quantum information science it plays a central role in error correction~\cite{bennett1996a,cory1998,schindler2011}, algorithm development~\cite{josza1994, gross2009, nielsen2010, yamasaki2018} and enabling quantum supremacy~\cite{boixo2018, neill2018, arute2019}; in condensed matter systems, it is a key resource for studying many-body localization~\cite{Basko2006, Nandkishore2015, Strkalj2021}, measurement-induced phase transitions~\cite{Ferguson2020, liy2018, Szyniszewski2019} and fermionic~\cite{Shapourian2017}, anyonic~\cite{Campagnano2012} and topological systems~\cite{Liu2022}, among others. This widespread presence of the effects of entanglement stems from its ability to imbue correlations which can spread across large system sizes via interactions, generating large-scale effects.

Central to the study of entanglement is its quantification and measurement~\cite{vidal2000}. This pursuit has given rise to two large, but currently distinct domains of Multipartite Entanglement (ME) and Symmetry Resolved Entanglement (SRE). The former extends the classical bipartite formulation of entanglement to capture non-local correlations which can arise when three or more subsystems are considered~\cite{amico2008}. It has been shown that ME can be sensitive to a wider range of entanglement classes~\cite{walter2013, maciazek2018} as compared to bipartite measures, making ME a useful tool to characterize a broader spectrum of entanglement present in a system. It has also found application in the study of symmetry-protected topological (SPT) phases~\cite{zhang2018,pezz2017}, where it was shown that ME is a necessary feature of these phases, and can act as an identifier of topological phases and phase transitions due to sharp changes in its scaling.

On the other hand, SRE concerns the decomposition of the total entanglement of the system into the contributions from various charge sectors which arise due to the presence of symmetries, as a consequence of N\"other's theorem. SRE has found relevance in the study of topological floquet order~\cite{azses2021}, free bosonic quantum many-body systems~\cite{pirmoradian2024}, noisy quantum devices~\cite{vitale2022}, operator entanglement~\cite{rath2023}, and non-Hermitian systems~\cite{fossati2023}, among others. In particular, we highlight the application of SRE in the study of SPT phases~\cite{azses2020a, azses2020b, azses2024, monkman2023}, where it has been noted to be a robust order parameter for SPT phases which is theoretically rigorous as well as experimentally accessible.

Given the usefulness of both ME and SRE paradigms in the study and classification of SPT phases, we expect that a unified application of the two approaches could lead to a finer, novel, and effective approach to study and classify these phases. In particular, it could lead to new information about where in charge space do global correlations which define SPT phases live. Despite rapid progress on both ME and SRE frontiers, there has been relatively little work done on the intersection of the two. Berthiere and Parez~\cite{berthiere2023} have previously investigated the symmetry resolution of the reflected entropy. This measure, although defined for a pair of physical subsystems, probes the multipartite structure of an enlarged Hilbert space constructed from a canonical purification which involves the two physical subsystems and two auxiliary \lq reflected\rq\,copies. In this respect, it may be considered the first work on multipartite symmetry resolution. 

In the present article, we aim to shed light on questions such as "Can a symmetry-resolved ME measure provide deeper insights into the characteristics of exotic phases such as SPT phases?", and "Does equipartition among charge sectors, which is a ubiquitous feature of bipartite symmetry resolution, extend to multipartite measures?". In particular, we provide a first step towards a general SR-ME paradigm: we perform explicitly the symmetry resolution of Meyer and Wallach's $Q$ measure of global entanglement~\cite{meyer2002}, which is a genuine ME measure. We show that for Haar random states, equipartition of $Q$ holds to all orders for systems of distinguishable particles hosting a locally acting symmetry,  and find explicit finite size corrections which follow a power law scaling with the number of local degrees of freedom. 

Our method provides new insights into the structure of ME under symmetries. Most notably, we find cases where all sector-wise contributions can vanish, and entanglement stems instead from the inter-sector coherences only. Further, we believe our methods and results yield themselves well to experimental verification, and we provide an explicit pipeline to extract individual sector contributions as well as those from sector coherences.

In order to ensure that the systems we are studying possess only genuine quantum correlations and have no thermal correlations, in this work we restrict our attention to pure quantum states, i.e. states described by a single vector in a complex vector space~\cite{wootters2001}. This will eliminate any possible classical correlations and ensure that the effects seen are purely due to quantum effects. Further, it is known to be an NP hard problem to determine whether a mixed state is even entangled or not~\cite{carisch2023, gurvits2003, gharibian2009}, and so for the sake of numerical tractability the restriction to pure states is crucial.

This paper is organized as follows: Sec.~\ref{sec:rev} is dedicated to a brief review of SRE and ME (in particular, Meyer and Wallach's $Q$ measure of global entanglement~\cite{meyer2002}). Then, in Sec.~\ref{sec:SR} we perform the symmetry resolution of $Q$. We support and verify the results obtained in this section using numerical simulations in Sec.~\ref{sec:numerics}. Further, we provide a detailed discussion of the relevant physical interpretations and broader implications of our findings in addition to opportunities for future research in Sec.~\ref{sec:disc}. Finally, we conclude in Sec.~\ref{sec:conc}. 

\section{Preliminaries}
\label{sec:rev}
\subsection{Symmetry Resolved Entanglement}
Consider a Hamiltonian $H$ which is invariant under the local action of a group $G$ via a representation $R$. Let the decomposition of $R$ into irreducible representations $R_\alpha$ take the form
\begin{equation}
\label{eq:rep_decomp}
    R = \bigoplus_\alpha R_\alpha ^{\chi_\alpha},
\end{equation}
where $\chi_\alpha$ are the multiplicities of the irreps $R_\alpha$. This in turn implies the decomposition of the local Hilbert space $\mathcal{H}$ as
\begin{equation}
\label{eq:space_decomp}
    \mathcal{H} = \bigoplus_\alpha \mathcal{H}_\alpha,
\end{equation}
with
\begin{equation}
\label{eq:dim_sum}
    d = \dim \mathcal{H} = \sum_\alpha \dim \mathcal{H}_\alpha = \sum_\alpha d_\alpha.
\end{equation}
It is easy to see that 
\begin{equation}
\label{eq:dalpha}
    d_\alpha = \chi_\alpha \dim R_\alpha, 
\end{equation}
where $\dim R_\alpha$ is the dimension of the irrep $R_\alpha$. SRE then asks what the contributions from each subspace $\mathcal{H}_\alpha$ are to the total entanglement. Operationally, SRE has been proposed to denote the entanglement remaining in the system after a measurement of charge on the system~\cite{goldstein2018, estienne2021} which causes the state to collapse into one of the subspaces $\mathcal{H}_\alpha$. In general, however, the sum of such charge-sector contributions to the entanglement is not necessarily equal to the total entanglement. Importantly, a lot of entanglement can arise from the interference between charge sectors rather then arising from the sectors themselves~\cite{carisch2023, macieszczak2019, ma2022}. 

SRE has been extensively investigated in 1-D spin chain models~\cite{goldstein2018, bonsignori2019, xavier2018, piroli2022, jones2022}. The majority of work has been for the the resolution of Abelian symmetries, while that of non-Abelian symmetries has only recently seen progress~\cite{bianchi2024, calabrese2021}. In this work, we do not impose Abelian nature as a requirement on our symmetry. Further, there exist experimental methods of measurement of SRE and its unique signatures. An explicit, step-by-step such method was presented in Ref.~\cite{neven2021}, and other experimental protocols were proposed and carried out in Refs.~\cite{azses2020b, vitale2022, rath2023}. 

A crucial and nearly ubiquitous feature of SRE is that of equipartition. At leading order, it has been shown analytically~\cite{xavier2018, kusuki2023, goldstein2018, bonsignori2019, jones2022} for finite/compact non-anomalous symmetries that the entanglement contribution of all charge sectors is the same. In particular, Goldstein and Sela~\cite{goldstein2018} showed in a seminal paper that in a $1 + 1$-D CFT for fermions with a $U(1)$ symmetry, at the quantum critical point, the $\ln L$ scaling of the entropy (in accordance with the violation of area laws~\cite{hastings2007, vidal2003, eisert2010}) is composed of $\sqrt{\ln L}$ contributions from charge sectors, $L$ being the subsystem size. Equipartition was interpreted in Ref.~\cite{bonsignori2019} as the entanglement contributions from each charge sectors being equal, simply with the probability of being in said charge sectors being unequal. However, note that there are next-to-leading order corrections~\cite{estienne2021, xavier2018, jones2022, ares2022} from effects such as finite system sizes~\cite{estienne2021, parez2021} and breaking of equipartition depending on the irreducible representation of the symmetry group being realized~\cite{kusuki2023, calabrese2021}. 

\subsection{Multipartite Entanglement and the \texorpdfstring{Q}{Q} measure}
\label{sec:Q}
On the other hand, ME concerns the quantification of the amount of entanglement in a system broken up into 3 or more subsystems. Further, ME measures are likely to be a better indicator of quantum phase transitions (QPTs), as opposed to conventional bipartite measures. In Ref.~\cite{osterloh2002} it was shown that the bipartite entanglement between two lattice sites is zero when the distance between them is larger than two spacings. However, at criticality, long-range order and entanglement is expected and a vanishing bipartite entanglement does not reflect this behaviour. In contrast, the entanglement sharing hypothesis~\cite{osborne2002} argues for the increase of ME at the expense of BE near criticality. Indeed, there is much evidence for this claim~\cite{li2024, iftikhar2021, giampaolo2013, montakhab2010, hofmann2014, sun2024}, providing further support to the need for ME measures in general.

We turn our attention to a specific ME measure, that introduced by Meyer and Wallach~\cite{meyer2002} based on the behaviour of states which are expected to show structure beyond simple bipartite entanglement. Dubbed the global entanglement, it is particularly conducive to symmetry resolution via a simple algebraic approach, as we demonstrate in Sec.~\ref{sec:SR}. Below we present the quantification of this ME measure as presented originally by Meyer and Wallach.

Denote pure states of an $n$-qubit system by $\ket{b_1 b_2 \ldots b_n}$ with $b_i \in \{0,1\}$ denoting the spin-up/down states of the qubit. Define a map $\iota_j(b)$ which acts on the state $\ket{b_1 b_2 \ldots b_n}$ by removing the $j^\mathrm{th}$ qubit if it matches $b \in \{0,1\}$, and annihilating it otherwise:
\begin{equation}
	\iota_j(b) \ket{b_1 b_2 \ldots b_n} = \delta_{bb_j}\ket*{b_1 b_2 \ldots \hat{b}_j \ldots b_n},
\end{equation}
where $\hat{b}_j$ represents omission. Next, define the squared norm of the wedge product between two vectors $u$ and $v$ as
\begin{equation}
	D(u,v) = \norm{u \wedge v}^2 = \sum_{x<y} \norm{u_x v_y - u_y v_x}^2,
\end{equation}
where $u_i, v_i$ are components of the vectors $u,v$. This is equal to $1 - F(u,v)$ where $F(u,v) = \norm{\bra{u} \ket{v}}^2$ is the fidelity of states $u$ and $v$~\cite{josza1994}. Finally, in terms of the map $\iota$ and the function $D(u,v)$, the global entanglement $Q$ of a pure state $\ket{\psi} = \ket{b_1 b_2 \ldots b_n}$ is given by
\begin{equation}
\label{eq:Q_original}
	Q(\ket{\psi}) = \frac{4}{n} \sum_j D(\iota_j(0) \ket{\psi}, \iota_j(1) \ket{\psi}).
\end{equation}
Physically, the $j^\mathrm{th}$ term in the sum describes how much the state of the system changes when the $j^\mathrm{th}$ qubit is removed. If the qubit removed was highly entangled with the others, then the system states resulting from removing $0$ and $1$ states of the qubit will differ significantly, and will possess higher orthogonality. Thus a low value of $Q$ corresponds to a system with low entanglement. The average over qubits ensures that the measure does not prefer a particular qubit, and that the measure is truly global. It accounts for the effect of all qubits, and the normalization factor of 4 ensures that $Q$ takes value in the unit interval. 

Brennen~\cite{brennen2003} showed that the global entanglement $Q$ has an alternative formulation: is equal to the average single-qubit linear entropy\footnote{We chose the normalization of the linear entropy such that it attains a maximum value of unity} of the system,
\begin{equation}
	\label{eq:Q_avg_lin_entr}
    Q = \langle S_L \rangle_k = 2 \langle 1 - \Tr(\rho_i^2) \rangle_k,
\end{equation}
where $\rho_k$ are single-qubit reduced density matrices (RDMs) and $\langle \cdot \rangle_k$ denotes an average over all qubits. This allows an easier interpretation to $Q$, namely the average entanglement of each qubit with the remainder of the system. This formulation also shows explicitly that $Q$ is an entanglement monotone. 

Further, $Q$ possesses several properties expected of an ME measure~\cite{meyer2002}: it takes on vanishing values on separable states, and maximal values on states such as the Bell states, the GHZ state, and the ground state of a spin-$\frac{1}{2}$ antiferromagnetic Ising chain which are expected to host maximal entanglement~\cite{xin2005, wang2010}. The $Q$ measure is also scalable, which allows for a general characterization of the entanglement in a system of qubits independent of the system size. In addition, it is also easily calculable, especially in its expression as the average linear entropy - yielding itself to easy interpretation as well as numerical analysis. This is further complemented by the fact that it is a physically relevant measure, and provides genuine physical insights into the system under consideration: it was shown in Ref.~\cite{deOliveira2006a} that $Q$ is a good indicator of phase transitions. In accordance with previous results on area laws~\cite{hastings2007, vidal2003,eisert2010,jin2004,keating2005,latorre2004,barthel2006,orus2018,tajik2023,herdman2017,lin2024,low2024}, $Q$ takes on a maximum value at the critical point for a spin-$\frac{1}{2}$ transverse field Ising chain, and in general, is as good as the linear and von Neumann entropies\footnote{This is expected to hold, at least, for all spin Hamiltonians with nearest-neighbour interactions~\cite{deChiara2018, wu2004} - although this can fail for systems with beyond nearest-neighbour interactions~\cite{yang2005}}.

Finally, in line with its physical relevance, we note that $Q$ is also experimentally accessible. Brennen~\cite{brennen2003} showed that the value of $Q$ for a system can be observed without requiring full quantum state tomography, instead using more efficient optical lattice techniques. In particular, it was shown that only two copies of the system are required to measure $Q$, while $N^2 -1 $ copies are needed for full quantum state tomography on an $N$-dimensional system. 

$Q$ can be generalized to systems composed of $d$-level particles, where the $d=2$ case reduces to qubits as discussed above. Since the linear entropy $1 - \tr(\rho^2)$ takes values from 0 to $\frac{d-1}{d}$ for systems living in $d$-dimensional Hilbert spaces~\cite{zyczkowski2006, pauletti2003}, the generalized definition of $Q$ becomes~\cite{scott2004}
\begin{equation}
\label{eq:qudit_Q}
   Q = \frac{d}{d-1} \Big \langle 1 - \Tr \rho_k^2\Big \rangle_k.
\end{equation}

In light of the above discussion, we believe that the global entanglement is a very interesting measure to study in the context of symmetry resolution. We now proceed to formally perform the symmetry resolution of the global entanglement $Q$.

\section{Symmetry resolution of \texorpdfstring{Q}{Q}}
\label{sec:SR}
Consider a lattice of arbitrary physical dimension consisting of particles with spin $s$, governed by a Hamiltonian which exhibits a symmetry under group $G$. Importantly, we require that the symmetry acts \textit{locally}, i.e. on each particle individually. If it is generated by an operator $X$ on a single particle, then the full symmetry acts as $\prod_i X_i$ where $i$ is an index over particle sites. We assume for now that each site hosts a single particle, and thus the Hilbert space per site $\mathcal{H}$ is $d$-dimensional with $d = 2s+1$. For a system of $n$ particles, we find that (see Appendix~\ref{app:a} for details) the sector-wise and interference contributions to the total entanglement for any system of distinguishable spins possessing a locally acting symmetry take the form
\begin{equation}
	\label{eq:final_intra}
	Q_\alpha = \frac{2d}{d-1}\Big\langle f_2^{(d)}(\rho_{k,\alpha}) \Big\rangle_k,
\end{equation}
and
\begin{equation}
	\label{eq:final_inter}
	Q_{\alpha\beta} =\frac{2d}{d-1} \Big\langle p_{k,\alpha}p_{k,\beta} +f_2^{(d)}(\rho_{k,\alpha \beta}) \Big\rangle_k,
\end{equation}
with 
\begin{equation}
\label{eq:final_result}
    Q = \sum_\alpha Q_\alpha + \sum_{\alpha < \beta} Q_{\alpha \beta}.
\end{equation}
Here $\rho_{k,\alpha} = P_\alpha \rho_k P_\alpha$ is the reduced density matrix (RDM) of the $k^\mathrm{th}$ particle after projecting onto the charge-$\alpha$ subspace using the corresponding projector $P_\alpha$, while $p_{k,\alpha} = \Tr(\rho_{k,\alpha})$ is the probability of the state to reside in the subspace . Further, $\rho_{k,\alpha\beta} = P_\alpha \rho_k P_\beta + P_\beta \rho_k P_\alpha$ is the interference contribution to the RDM of the $k^{\mathrm{th}}$ particle, $\langle \cdot \rangle_k$ represents an average over particles, and $f_2^{(d)}(A)$ is the sum of all pairwise products of eigenvalues of $A$:
\begin{equation}
\label{eq:f2d}
    f_2^{(d)}(A) = \sum_{i < j} \lambda_i\lambda_j = \frac{(\Tr A)^2 - \Tr (A^2)}{2}.
\end{equation}
Equivalently, it is the coefficient of $\lambda^{d-2}$ in the characteristic polynomial of $A$ or the sum of all principal $2 \times 2$ minors of $A$. We now discuss each of the two terms (Eqs.~\eqref{eq:final_intra} and~\eqref{eq:final_inter}) in more detail.

\subsection{Sector-wise contributions}
\label{sec:sector}

Since the sector-wise contributions in Eq.~\eqref{eq:final_intra} depend solely on $f_2^{(d)}(\rho_{k,\alpha})$, we see from Eq.~\eqref{eq:f2d} that the quantities of interest are the eigenvalues of $\rho_{k,\alpha}$. In particular, note that $f_2^{(d)}(\rho_{k,\alpha})$ - and thus the sector-wise contributions - must vanish completely when $\mathrm{rk}(\rho_{k,\alpha}) \leq 1$, where $\mathrm{rk}$ denotes the rank of the matrix. Using the property
\begin{equation}
	\label{eq:rank_rel}
	\mathrm{rk}(AB) \leq \min(\mathrm{rk}(A), \mathrm{rk}(B)),
\end{equation}
we see that $\mathrm{rk} (\rho_{k,\alpha}) \leq \min (\mathrm{rk}(P_\alpha), \mathrm{rk}(\rho_k))$. There are then two cases\footnote{Note that the case $\mathrm{rk}(\rho_k) = 0$ for all $k$ is impossible due to the unit-trace constraint on density matrices, and the $\mathrm{rk}(P_\alpha) = 0$ case is trivial since it implies that representation corresponding to charge $\alpha$ is entirely absent from the decomposition~\eqref{eq:rep_decomp}} where intra-sector contributions to $Q$ can vanish: $\mathrm{rk}(\rho_k) = 1$ for all $k$, or $\mathrm{rk}(P_\alpha) = 1$.

Since the $\rho_k$ eigenvalues represent probabilities, $\mathrm{rk} (\rho_{k}) = 1$ denotes that the particle at site $k$ has only one possible state $\ket{\psi_k}$. If all particles are in such a state, then the state of the full system can then be factorized as $\ket{\psi} = \bigotimes_k \ket{\psi_k}$, i.e., the system is in a fully separable (non-entangled) state. The global entanglement $Q$ (and indeed, all entanglement monotones) vanishes on such states, which explains the vanishing sector-wise contributions.

The other case, $\mathrm{rk}(P_\alpha) = 1$, is more interesting. Noting that the rank of the projector is equal to the dimension of the subspace it projects into, $\mathrm{rk} (P_\alpha) = d_\alpha$, the only possible positive integer values consistent with Eq.~\eqref{eq:dalpha} are $\dim(R_\alpha) = \chi_\alpha = 1$, which implies that $P_\alpha$ can have unit rank if and only if the corresponding irrep $R_\alpha$ is one-dimensional and occurs with multiplicity $\chi_\alpha = 1$ in the action of the group $G$. This implies that the contributions $Q_\alpha$ in Eq.~\eqref{eq:final_intra} to the global entanglement $Q$ from charge sectors $\alpha$ vanish if and only if the charge $\alpha$ subspace has dimension $1$ (equivalently, the dimension $\dim(R_\alpha)$ and the multiplicity $\chi_\alpha$ in $R$ are both unity).

This rather surprising result can be understood in terms of the post-charge-measurement interpretation of SRE. If the state is projected onto a one-dimensional subspace, it is specified fully (up to a global phase) and all information about it is known - leading to vanishing entropy and thus no entanglement contribution from that sector. This is, however, not true for charge subspaces of dimension larger than one, since projection onto  such subspaces fails to fully specify the state - generically allowing for entanglement to reside in larger subspaces.

The above result can be combined with the fact that all complex irreps of Abelian groups are one-dimensional (a consequence of Schur's Lemma), and that the regular representation takes the form
\begin{equation}
    R_\mathrm{reg} = \bigoplus_\alpha R_\alpha^{\dim R_\alpha}.
\end{equation}
In other words, for a finite Abelian group acting on a system via its regular representation, all irreps $R_\alpha$ have $\dim (R_\alpha) = \chi_\alpha = 1$. This immediately leads to the corollary that for a finite Abelian symmetry group $G$ acting locally via its regular representation on a system of distinguishable spins, \textit{all} single-sector contributions to the global entanglement $Q$ vanish. Consequently, entanglement stems solely from the pairwise interference between sectors and is given by Eq.~\eqref{eq:final_inter}. We will verify this numerically in Sec.~\ref{sec:numerics}. 

Note further that there is another case where $f_2^{(d)}$ must vanish for all particles $k$. When $d=2$, $f_2^{(d)}$ is simply the determinant. In this case, then
\begin{equation}
\label{eq:det_rho_k_alpha}
	f_2^{(d)}(\rho_{k,\alpha}) = \det (\rho_{k,\alpha}) = \det(P_\alpha)\det(\rho_k)\det(P_\alpha).
\end{equation}
As long as there is more than one charge sector, the projectors $P_\alpha$ must project into subspaces $\mathcal{H}_\alpha$ of dimension strictly lower than the dimension of the full single-site Hilbert space $\mathcal{H}$. This means their determinant must necessarily vanish, in turn implying vanishing values of $f_2^{(d)}(\rho_{k,\alpha})$ by Eq.~\eqref{eq:det_rho_k_alpha}, as well as vanishing sector-wise contributions by Eq.~\eqref{eq:final_intra}. Alternatively, this can be seen via Eq.~\eqref{eq:dim_sum}, where the only non-trivial (more than one charge sector) partition of $d = 2$ is $2 = 1 + 1$, and this restricts the possible local group actions on qubits to have $\dim(R_\alpha) = \chi_\alpha = 1$ with only two possible non-trivial charge sectors $\alpha, \beta$. We thus see that for systems of distinguishable spin-$\frac{1}{2}$ particles (qubits), all possible local symmetry group actions with more than one symmetry sector lead to vanishing contributions $Q_\alpha$ to the total global entanglement $Q$.

In Appendix~\ref{app:b} we explicitly calculate the individual sector contributions, and show that equipartition holds to all orders, showing the leading order behaviour and all relevant corrections. We find that for a system of $n$ particles of local Hilbert space dimension $d$, repeatedly realized in a pure state chosen randomly over the Bloch hypersphere (Haar random states), the individual sector contributions take an average of
\begin{equation}
\label{eq:equipartition_intra}
    Q_\alpha = \frac{d_\alpha(d_\alpha-1)}{d(d-1)} \;\bigg( \frac{1-d^{1-n}}{1+d^{-n}} \bigg).
\end{equation}

We see that the contributions depend solely on the dimensions of the local Hilbert space and the charge subspace up to leading order, and that it is independent of the symmetry group acting on the system. Further, the sub-leading order corrections can be interpreted as finite size corrections, since they carry all $n$-dependence and vanish in the large $n$ limit.

We recall the discussion of equipartition under symmetry resolution from Sec.~\ref{sec:intro}. Therein it was noted that Bonsignori \textit{et al}.~\cite{bonsignori2019} interpreted equipartition as equal contributions from each sector, but with the probability of a state to be in each sector being different. In other words, the contribution from a charge sector is the product of its intrinsic contribution and the probability of the state being in the charge-$\alpha$ sector. However, we have shown earlier that it is impossible for a one-dimensional charge subspace to host entanglement, and that $d_\alpha \geq 2$ is necessary. In this light, we reinterpret equipartition for the global entanglement $Q$ to mean instead equal contributions from each charge sector, but with the probability of \textit{entanglement being hosted} in each sector being different (contrast with Bonsignori \textit{et al.}'s interpretation). This is manifest in Eq.~\eqref{eq:equipartition_intra} if we write $Q_\alpha$ to leading order as ${d_\alpha \choose 2}/{d \choose 2}$: the sector-wise contributions are precisely the probability of two randomly chosen basis vectors to lie in the same charge sector of dimension $d_\alpha$. We thus conclude that the intrinsic contribution of each charge sector is the same, and that Eq.~\eqref{eq:equipartition_intra} explicitly conforms to the equipartition expectation from a symmetry-resolved entanglement measure - albeit with the slightly modified interpretation of equipartition.

In essence, we have related the charge-sector contributions to the (ordered) integer partition of the local Hilbert space dimension, where each summand denotes the dimension of a charge subspace. Below, we see how this extends to the contributions from the interference between charge sectors.

\subsection{Interference contributions}
\label{sec:interference}

Using analysis similar to that for the sector-wise terms, in Appendix~\ref{app:b} we calculate the ensemble average of the interference contributions for a set of states chosen uniformly over the Bloch sphere. We find a form very similar to the sector-wise contributions:
\begin{equation}
\label{eq:equipartition_inter}
    Q_{\alpha\beta} = \frac{2 d_\alpha d_\beta}{d(d-1)} \:\bigg( \frac{1 - d^{1-n}}{1+d^{-n}}\bigg).
\end{equation}
Interestingly, the finite size correction factor is the same as that for the sector-wise contributions, see Eq.~\eqref{eq:equipartition_intra}. Further, the leading order behaviour conforms once again to the modified interpretation of equipartition: there are precisely $d_\alpha d_\beta$ ways to choose two basis vectors, one from each of the $\alpha$ and $\beta$ subspaces. The probability is then $d_\alpha d_\beta/{d\choose2}$, yielding the leading order behaviour of the interference contributions.

Another interesting observation is that interference contributions (and indeed, the sector-contributions as well) arise from the choice of only two basis vectors in the local Hilbert space. Algebraically, this arises due to the quadratic $\rho$-dependence of the purity, and consequently of the global entanglement. Physically, this seems to suggest that the global entanglement is insensitive to any fundamental three-way entanglement present in nature (i.e. which cannot be written in terms of the substituent two-way entanglements), if there exists any. 

In all, this indicates equipartition not only among the charge sector contributions, but in the full set of charge-sector and interference contributions. Each contributor has an intrinsic value of $2/d (d-1)$ to leading order, with entanglement being hosted in them in $d_\alpha (d_\alpha-1)/2$ and $d_\alpha d_\beta$ ways for sector-wise and interference contributions respectively. This completes extraction of symmetry-resolved contributions to the global entanglement $Q$. We now proceed to verify our results numerically.

\section{Numerical tests}
\label{sec:numerics}
In this section, we perform numerics on ensembles of Haar random states. These are states sampled over a uniform distribution on the surface of the Bloch hypersphere, ensuring that they are indeed pure states. They represent benchmark states which are known to saturate entanglement measures in the bipartite case~\cite{page1993}. While they are not representative of low energy states or ground states of systems, they provide a useful comparison as a maximally entangled null model. Haar random states have also been used previously in the study of SRE~\cite{murciano2022, ghasemi2025}, since they isolate the effect of symmetry on the system from other non-trivial physical effects such as topological order, boundary modes, etc. This can also be used to compare against systems which possess such characteristics, and deviations from Haar ensemble behaviour can lead to insights on SPT phases~\cite{azses2020a, azses2024, monkman2023} for example, where the added effect of topology can be identified with the symmetry-only effect in the Haar model. 

For different values of the tuple $(n,d)$, sets of 500 $n$-particle states were generated by randomly selecting states $\ket{\psi} = \sum_i^{d^n} q_i \ket{u_i}$ with $\{\ket{u_i}\}$ being the $n$-particle basis states. The coefficients $q_i \in \mathbb{C}$ were selected uniformly over the Bloch hypersphere, ensuring a normalized pure state as output. The density matrices and single-particle RDMs for all states were found simply by implementing the outer product and partial trace operations successively.  These objects then allowed us to obtain the individual values of all the terms in Eqs.~\eqref{eq:final_intra} -~\eqref{eq:final_inter} for each of the samples: $f_2^{(d)}(\rho_{k,\alpha})$ for the sector contributions, and $\Tr(\rho_k P_\alpha) \Tr(\rho_k P_\beta)$ and $f_2^{(m)}(\rho_{k,\alpha \beta})$ for the interference contributions. 

To fingerprint our results, we implemented an Abelian symmetry on the generated Haar random states. In particular, we imposed a $\mathbb{Z}_m$ symmetry acting on each particle via a representation as defined in Eq.~\eqref{eq:dim_sum}, with all irreps having dimension $\dim R_\alpha=1$ and multiplicities obeying $\sum_{\alpha=1}^m \chi_\alpha = d$. In other words, the ordered partition $d = \sum_\alpha d_\alpha$ with $d_\alpha = \chi_\alpha$ fully determines the action of the symmetry group on the system up to relabelling of basis vectors.

\begin{figure}[t!]
    \centering
    \includegraphics[width=\linewidth]{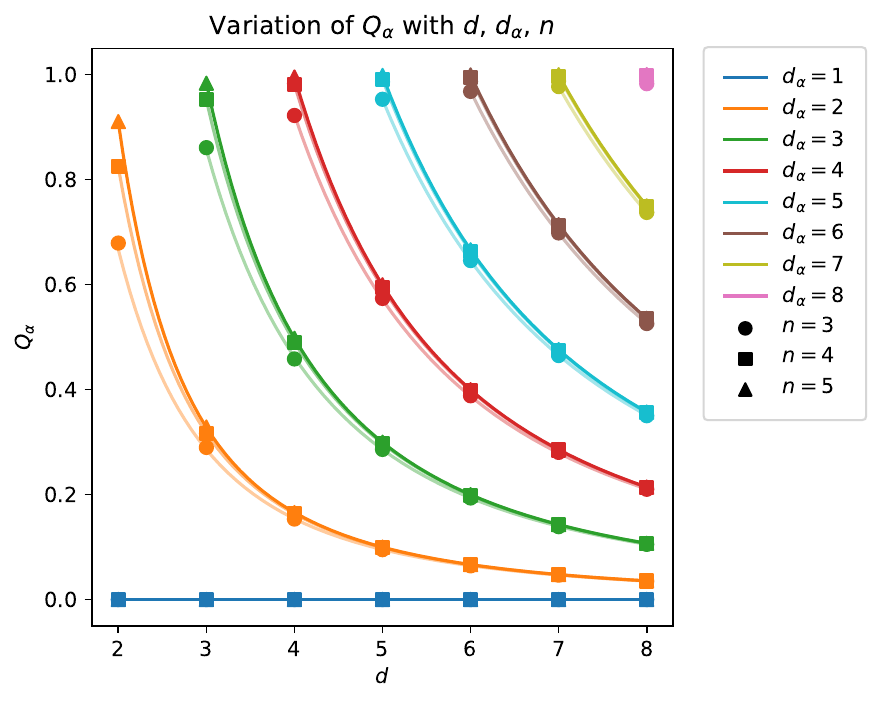}
    \caption{Scatter plot of $Q_\alpha$ versus $d$, with the prediction of Eq.~\eqref{eq:final_intra} laid out in coloured lines. Each set of three closely spaced lines corresponds to a single value of $d_\alpha$ with varying $n$. While $Q_\alpha$ vanishes for all values of $n$ and $d$ when $d_\alpha$ = 1, higher values of $d_\alpha$ generically possess higher entanglement for a given $d$. Note also that $Q_\alpha \neq 0$ for $d_\alpha=d=2$ arises since there is only one charge sector in that case, and so does not contradict the prediction of vanishing sector-contributions for qubit systems with symmetries possessing multiple charge sectors. Further, larger system sizes $n$ possess higher entanglement, although successive corrections become smaller with increasing $n$. Overall, data shows excellent agreement to the prediction, with standard deviation of $\sim10^{-3}$.}
    \label{fig:sector}
\end{figure}

We first verify the equivalence of three distinct ways of calculating $Q$: it's original formulation as presented by Meyer and Wallach~\cite{meyer2002}, its quantification as the average single-particle linear entropy~\cite{brennen2003} and the symmetry-resolved decomposition presented herein. We find perfect numerical agreement between the $Q$ values for all tested values of $m $, verifying our analytics in Sec.~\ref{sec:SR}. Next, we verify our claim of vanishing sector-wise contributions for $\chi_\alpha = 0, 1$ and the dependence on general $d_\alpha$ as quantified in Eq.~\eqref{eq:final_intra}. For $n=3,4,5$ and for various values of $d_\alpha$, we plot the ensemble average of $Q_\alpha$ as a function of $d$ in Fig.~\ref{fig:sector}. We find strong agreement with the prediction of Eq.~\eqref{eq:final_intra}, when taking into account the finite size corrections. We re-emphasize the independence of the results on $m$, and the vanishing contributions for $d_\alpha = 1$. 

Next, we simulate the behaviour of the interference term. As expected from Eq.~\eqref{eq:final_inter}, it follows the same $d$ and $n$ dependence as the sector-wise contributions. We find that the numerical results and the expectation from Eq.~\eqref{eq:equipartition_inter} agree to one part in $10^{-3}$. We refrain from plotting a figure in light of the similarity to Fig.~\ref{fig:sector}.

\section{Discussion}
\label{sec:disc}
The above numerical simulations verify our analytical treatment in Sec.~\ref{sec:SR}. In particular, they lend strong support to the post-charge-measurement interpretation of SRE and the reinterpretation of equipartition in terms of the probabilities of entanglement being hosted in a particular charge sector. Overall, our treatment conforms to expectations from the symmetry resolution of a multipartite entanglement measure without locality, topology or other non-trivial physical effects, namely equipartition and lack of dependence on a particular subsystem. 

We stress further that the surprising vanishing of sector-wise contributions in some cases is a feature not observed in bipartite SRE. The dependence on the particular subsystem in bipartite scenarios implies that the entanglement measured does not reflect the properties of the entire system. The restriction of the charge sector to the particular subsystem is blind to the global interference within the charge sector which could potentially cancel out all contributions from the sector. 

An important assumption in our work is that each lattice site hosts only a single particle. If we were to coarse-grain and allow multiple particles per site, the calculation above fails, owing to the additional quantum number required to fully describe the system state. The added degree of freedom changes the Hilbert space dimension per site, and ignoring it leads to lower information about the system. In turn, this leads to higher entropy and thus entanglement, a significant part of which can now be hosted within a single site.

The other assumption made above was that of a locally acting symmetry. In addition to allowing definition of local projection operators $P_\alpha$, it ensures that the RDMs possess a block‑diagonal structure in the symmetry basis. Locally acting symmetries are also conducive to experiments, since the observables associated to such symmetries often correspond to local spin and density operators. These, in turn, are directly measurable in quantum simulators. Such measurements have been performed in ultracold atoms using site- and spin-resolved probes of magnon dynamics~\cite{fukuhara2013} and via quantum gas microscopy~\cite{christakis2023}. Such measurements have also been performed in ion traps, fluorescence detection in out-of equilibrium spin chains~\cite{friedenauer2008,neyenhuis2017}.

Further, we believe the results of this work are experimentally verifiable. As mentioned in Sec.~\ref{sec:Q}, the global entanglement $Q$ is proposed to be measurable with only two copies of the state~\cite{brennen2003}. Combined with the post-charge-measurement interpretation of SRE, this is likely the easiest way to reconstruct the single-sector terms. Explicitly, we expect charge measurements on a pure state followed by measurements of $Q$ on the collapsed state to yield the contributions $Q_\alpha$. On the other hand, the value of $\langle f_2^{(d)} (\rho_{k,\alpha})\rangle_k$ can be estimated using the expression of $f_2^{(d)}(A)$ in terms of the traces $\Tr A$ and $\Tr (A^2)$. This can be achieved, for example, using SWAP-gate techniques introduced in Ref.~\cite{ekert2002} and performed, for example, in~\cite{islam2015}, which allow estimation of linear and non-linear functionals of the RDM without full quantum state tomography. Randomized measurements~\cite{brydges2019} also holds promise as possible experimental methods to estimate $f_2^{(d)}(\rho_{k,\alpha})$. A comparison of the measured $Q_\alpha$ and $\langle f_2^{(d)}(\rho_{k,\alpha})\rangle_k$ would then be a straightforward experimental test of Eq.~\eqref{eq:final_intra}.

\begin{figure}
    \centering
    \includegraphics[width=\linewidth]{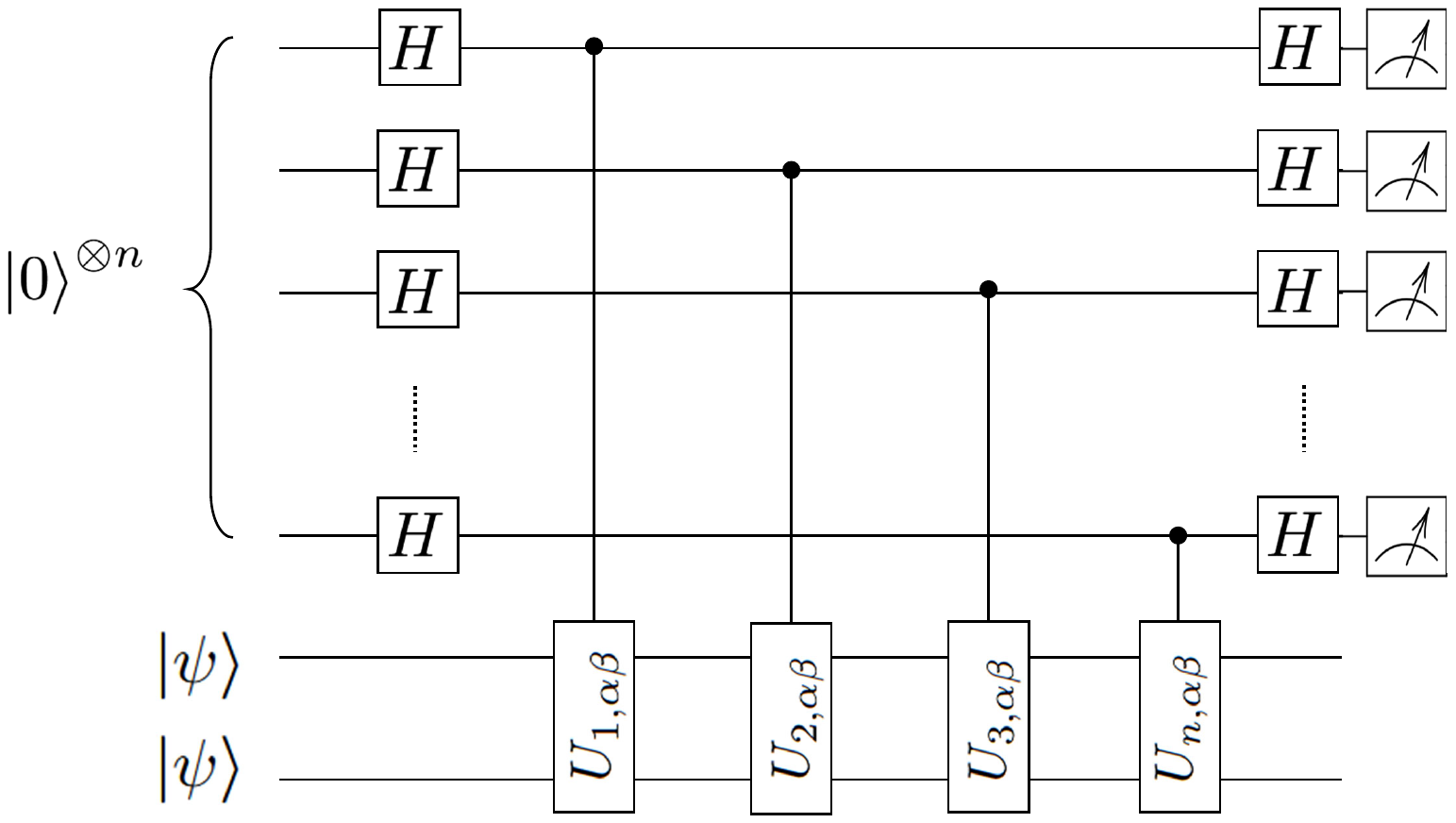}
    \caption{Quantum circuit showing the proposed measurement technique for the interference contributions. The ancillary register, prepared initially in a state $\ket{0}^{\otimes n}$ is subjected to repeated controlled-$U_{k,\alpha\beta}$ gates, one for each particle in the system $\ket{\psi}$. Such gates entangle the $k^\mathrm{th}$ ancillary qubit with two copies of the $k^\mathrm{th}$ system particle in the $\alpha$ and $\beta$ charge sectors. Subsequent measurement of the ancillary register for multiple such experiments allows a full reconstruction of $\Tr(\rho_k P_\alpha \rho_k P_\beta)$, in turn leading to the reconstruction of the full interference term in Eq.~\eqref{eq:final_inter} when combined with measurement of $p_{k,\alpha}$. }
    \label{fig:circuit}
\end{figure}

As for the interference contributions, a charge measurement will fail since it annihilates all the interference contributions. However, we detail in Appendix~\ref{app:c} the theory behind a possible experimental verification of Eq.~\eqref{eq:final_inter}, showing that full reconstruction of $Q_{\alpha \beta}$ is still possible. To do this, we utilize projective measurements to isolate particular interference contributions, and a controlled-unitary quantum circuit (see Fig.~\ref{fig:circuit}) to reconstruct $f_2^{(d)}(\rho_{k,\alpha\beta})$. The full working of the circuit is presented in Appendix~\ref{app:c}.

\section{Conclusion}
\label{sec:conc}
In this work we demonstrated explicitly the symmetry resolution of the global entanglement measure $Q$ under the action of local symmetries on spin lattice systems of arbitrary particle spin and physical dimension. We characterized the contributions from each charge sector (Eq.~\eqref{eq:final_intra}) and from their pairwise interference (Eq.~\eqref{eq:final_inter}) by transcribing the action of the symmetry group on the local Hilbert space to an ordered partition of its total dimension. Importantly, we note that the sector-wise contributions vanish whenever the subspace corresponding to the charge sector is one-dimensional. We thus conclude that the global entanglement can only be hosted in two- or higher-dimensional subspaces. We provide further intuition of this result within the post-charge-measurement interpretation of symmetry resolved entanglement. 

In addition, we find that equipartition holds for Haar random states of the entire system of $n$ particles, where we reinterpret equipartition to mean equal intrinsic contributions with differing probabilities of entanglement being hosted in each sector. This is a modification of a similar interpretation proposed previously by Bonsignori \textit{et al.}~\cite{bonsignori2019}. In particular, we find that in the limit of large system sizes, the contribution from each charge sector or their pairwise interference is equal to the probability of two chosen basis vectors to lie within the subspace under consideration (see Eqs.~\eqref{eq:equipartition_intra} and Eq.~\eqref{eq:equipartition_inter}). Moreover, we find arbitrarily accurate finite size corrections which take the same form for both types of contributions. We verify all our claims numerically and find perfect correspondence between theory and numerics. 

Finally, we propose experimental methods to verify our results using the post-charge-measurement interpretation of SRE. For the interference contributions (which are destroyed upon charge measurement on the system), we propose a quantum circuit which could be implemented to extract part of the interference contributions which can then be implemented to reconstruct the full interference contributions from each pair of charges.

In comparison to symmetry-resolved bipartite entanglement measures and symmetry-unresolved ME measures, SR-ME measures could provide access to knowledge about which sectors carry global coherence in SPT phases. In other words, it can shed light on the pattern of global coherences conditioned on charge, which could lead to a more nuanced classification of SPT phases whose non-trivial characteristics are non-local but charge-dependent.

\section*{Acknowledgements}
A large portion of this work was completed at the University of Oxford as part of the author's Masters Dissertation on a related topic. This was supervised by Dr. Nick Jones, to whom the author is grateful. The author also acknowledges enlightening discussions with Param Luhadiya, Vansh Jhunjhunwala, Yaprak \"Onder, Om Gupta, Aryaman Babbar and Lucas Gathercole. 

\begin{appendix}
\numberwithin{equation}{section}

\section{Derivation of sector-wise and interference contributions}
\label{app:a}
In this section we perform explicitly the symmetry resolution of the global entanglement $Q$. As in the main text, consider a Hamiltonian $H$ which describes a system of particles of spin-$s$, invariant under the local action of group $G$ via representation $R$ which decomposes into irreps $R_\alpha$ with multiplicities $\chi_\alpha$ according to Eq.~\eqref{eq:rep_decomp}. For each charge value $\alpha$, the spectral theorem guarantees the existence of single-particle charge subspace projectors $P_\alpha$, i.e. maps $P_\alpha:\mathcal{H} \rightarrow \mathcal{H}_\alpha$ which obey the following properties:
\begin{align}
	P_\alpha P_\beta = P_\beta P_\alpha &= 0 \;\;\;\;\mathrm{if}\,\alpha \neq \beta,\tag{Orthogonality} \\
	P_\alpha^2 &= P_\alpha, \tag{Idempotence} \\
	\sum_\alpha P_\alpha &= \mathbb{I}_m. \tag{Completeness}
\end{align}
For a system in a pure state $\ket{\psi}$, consider the density matrix $\rho = \ket{\psi}\bra{\psi}$, and trace out all but  the $k^\mathrm{th}$ particle to obtain the single-particle ($d \times d$) RDMs $\rho_k$. Now decompose these into sectors using two insertions of the identity and completeness:
\begin{align}
	\label{eq:rho_decomp}
	\rho_k &= \mathbb{I}_2\;\rho_k\;\mathbb{I}_2 \nonumber \\
	&= \bigg(\sum_\alpha P_\alpha\bigg) \; \rho_k \;\bigg(\sum_\beta P_\beta\bigg) \nonumber \\
	&=  \sum_\alpha P_\alpha \rho_k P_\alpha + \sum_{\alpha < \beta} P_\alpha \rho_k P_\beta + P_\beta \rho_k P_\alpha\nonumber \\
	&:= \sum_\alpha\rho_{k,\alpha} + \sum_{\alpha < \beta} \rho_{k,\alpha\beta}^\mathrm{int}.
\end{align}
Here we have defined the sector-wise components of the RDMs $\rho_{k,\alpha}$ and the cross-sector interference terms $\rho_{k,\alpha\beta}^\mathrm{int}$. Note that in the interference term, each pair $(\alpha, \beta)$ is counted only once since the sum is over $\alpha < \beta$, but the term is symmetric due to the symmetric definition of $\rho_{k,\alpha\beta}^\mathrm{int}$. We will drop the superscript \lq$\mathrm{int}$\rq~hereafter, assuming a two-index $\rho_{k,\alpha \beta}$ implicitly refers to the interference contributions. Recalling now from Eq.~\eqref{eq:qudit_Q} that $Q$ depends on the purities $\Tr \rho_k^2$, we can use Eq.~\eqref{eq:rho_decomp} to write
\begin{equation}
\label{eq:tr_expand}
	\Tr \rho_k^2 = \Tr \bigg( \sum_\alpha \rho_{k,\alpha} \bigg)^2 + \Tr \bigg(\sum_{\alpha < \beta}\rho_{k, \alpha \beta}\bigg)^2 + \Tr\bigg( \sum_{\gamma, \alpha < \beta} \{\rho_{k,\gamma}, \rho_{k,\alpha\beta}\} \bigg),    
\end{equation}
with $\{a,b\} := ab + ba$ being the anticommutator. In the final term, the sum can be pulled out of the trace by linearity.  Then all anticommutators traces are of the form
\begin{align*}
	\Tr \{\rho_{k,\gamma}, \rho_{k,\alpha \beta}\} &=  \Tr (\rho_{k,\gamma}\rho_{k,\alpha \beta}) + \Tr(\rho_{k,\alpha \beta} \rho_{k,\gamma}) \\
	&= \Tr(P_\gamma \rho_k P_\gamma P_\alpha \rho_k P_\beta) + \Tr(P_\gamma \rho_k P_\gamma P_\beta \rho_k P_\alpha) + \Tr(P_\alpha \rho_k P_\beta P_\gamma \rho_k P_\gamma) + \Tr(P_\beta \rho_k P_\alpha P_\gamma \rho_k P_\gamma),
\end{align*}
all of which must vanish by orthogonality of the projectors and the cyclicity of the trace, since $\alpha < \beta$ and $\gamma$ can be equal to only one of $\alpha, \beta$, if any. Further, note that the first term in Eq.~\eqref{eq:tr_expand} simplifies as
\begin{equation*}
	\Tr\bigg(\sum_\alpha \rho_{k,\alpha} \bigg)^2 =  \Tr\sum_\alpha \rho_{k,\alpha}^2 ,
\end{equation*}
since the cross-terms vanish by orthogonality of the projectors. Similarly, the second term can be written as
\begin{equation*}
	\Tr \bigg(\sum_{\alpha < \beta}\rho_{k, \alpha \beta}\bigg)^2 = \Tr\sum_{\alpha < \beta}\rho_{k, \alpha \beta}^2 
\end{equation*}
by orthogonality and the cyclicity of the trace. While it is possible to simplify further the expression $\rho_{k,\alpha\beta}^2$ by expanding, we keep it in this form for later simplification and analogy to the single-sector purities. Finally, this yields a simple relation between the purities
\begin{equation}
	\Tr \rho_k^2 = \sum_\alpha \Tr \rho_{k,\alpha}^2 + \sum_{\alpha < \beta} \Tr \rho_{k,\alpha \beta}^2.
\end{equation}
Using this, we can now write $Q$ from Eq.~\eqref{eq:qudit_Q} as 
\begin{align*}
	Q &:= \frac{d}{d-1}\Big\langle 1 - \Tr\rho_k^2\Big\rangle_k \\
	&= \frac{d}{d-1}\Big\langle1 - \sum_\alpha \Tr \rho_{k,\alpha}^2 - \sum_{\alpha < \beta} \Tr \rho_{k,\alpha \beta}^2 \Big\rangle_k,
\end{align*}
where the normalization is based on the local Hilbert space dimension. For our final trick, we note that upon measurement of the charge, the particle at site $k$ can collapse to a charge subspace $\alpha$ with probability $p_{k,\alpha}$. These probabilities are simply the expectation values of the projectors
\begin{equation}
\label{eq:prob_trace}
p_{k,\alpha} = \Tr(P_\alpha\rho_k) = \Tr(P_\alpha^2\rho_k) = \Tr(P_\alpha\rho_k P_\alpha) = \Tr(\rho_{k,\alpha}).     
\end{equation}
Further, these probabilities must sum to unity, i.e. $1 = \sum_\alpha p_\alpha$. On squaring, this yields $1 = \sum_\alpha p_{k,\alpha}^2 + 2\sum_{\alpha < \beta}p_{k,\alpha}p_{k,\beta}$. This can also be seen from taking the trace of Eq.~\eqref{eq:rho_decomp}, noting that $\Tr \rho_{k,\alpha\beta}$ vanishes by linearity and cyclicity of the trace. We now substitute this expression for unity in our expression for $Q$ to obtain
\begin{align}
	\label{eq:Q_exp}
	Q &= \frac{d}{d-1}\Big\langle \sum_\alpha p_{k,\alpha}^2 + 2\sum_{\alpha < \beta}p_{k,\alpha}p_{k,\beta} - \sum_\alpha \Tr \rho_{k,\alpha}^2 - \sum_{\alpha < \beta} \Tr \rho_{k,\alpha \beta}^2 \Big\rangle_k\nonumber  \\
	&= \frac{d}{d-1}\Big\langle \sum_\alpha \bigg( p_{k,\alpha}^2 - \Tr \rho_{k,\alpha}^2 \bigg)\Big\rangle_k + \frac{d}{d-1} \Big\langle \sum_{\alpha < \beta} \bigg( 2p_{k,\alpha}p_{k,\beta} - \Tr \rho_{k,\alpha \beta}^2 \bigg)\Big\rangle_k.
\end{align}
Here we have naturally regrouped terms by their symmetry sectors, and we can identify each term in Eq.~\eqref{eq:Q_exp} as a contribution from a symmetry sector or from their combined interference~\cite{castro2024}. Note that this was the initial aim of our symmetry resolution, and we have isolated the required contributions. We further simplify this by using Eq.~\eqref{eq:f2d}, noting simultaneously that $\Tr(\rho_{k,\alpha\beta}) = \Tr(P_\alpha \rho_k P_\beta + P_\beta \rho_k P_\alpha) = 0$. This allows us to write $Q$ from Eq.~\eqref{eq:Q_exp} as
\begin{equation}
	Q = \frac{2d}{d-1}\Big\langle \sum_\alpha \bigg(f_2^{(d)}(\rho_{k,\alpha}) \bigg)\Big\rangle_k + \frac{2d}{d-1} \Big\langle \sum_{\alpha < \beta} \bigg(p_{k,\alpha}p_{k,\beta} +f_2^{(d)}(\rho_{k,\alpha \beta})\bigg)\Big\rangle_k.
\end{equation}
This completes the derivation of Eqs.~\eqref{eq:final_inter} - \eqref{eq:final_result} in the main text.

\section{Equipartition and estimation of contributions for Haar random states}
\label{app:b}
Since each of the $n$ sites has a $d$-dimensional Hilbert space, there are $N = d^n$ basis states of the full system's Hilbert space. Denote the charge eigenbasis of this space by $\{\ket{v_i}\}$ with $i \in \{1,2\ldots N\}$. Let the system be in a state $\ket{\psi} = \sum_{i=1}^N q_i\ket{v_i}$ where $q_i \in \mathbb{C}$ obey $\sum_{i=1}^N |q_i|^2 = 1$. Here and for the numerical simulations in Sec~\ref{sec:numerics}, we take these $q_i$ to be sampled from a uniform distribution over the complex unit sphere with mean zero, i.e. a Haar random distribution. This is natural for a randomly selected state of the system. 

In order to calculate $Q_\alpha = \frac{2d}{d-1} f_2^{(d)}(\rho_{k,\alpha})$, we write
\begin{equation}
    \langle f_2^{(d)}(\rho_{k,\alpha}) \rangle_k= \bigg \langle \sum_{i < j} \mu_i \mu_j\bigg \rangle_k = \frac{1}{2} \bigg ( (\Tr \rho_{k,\alpha})^2 -  \Tr (\rho_{k,\alpha}^2)\bigg ),
\end{equation}
where $\mu_i$ are the eigenvalues of $\rho_{k,\alpha}$. Further, note that
\begin{equation}
    (\Tr\rho_{k,\alpha})^2 = \bigg (\sum_{i=1}^{d_\alpha} \sigma_{ii}\bigg )^ 2 = \sum_{i=1}^{d_\alpha} \sigma^2_{ii} +\sum_{i\neq j = 1}^{d_\alpha} \sigma_{ii} \sigma_{jj},
\end{equation}
where we denote $\rho_k$ by $\sigma$ for brevity of notation. Similarly, we can write the other trace in index notation as
\begin{equation}
    \Tr (\rho_{k,\alpha}^2) = \sum_{i, j = 1}^{d_\alpha} |\sigma_{i,j}|^2 =  \sum_{i=1}^{d_\alpha} |\sigma_{ii}|^2 +\sum_{i\neq j = 1}^{d_\alpha} |\sigma_{ij}|^2,
\end{equation}
allowing us to express 
\begin{equation}
    \langle f_2^{(d)}(\rho_{k,\alpha}) \rangle_k = \frac{1}{2} \bigg \langle \sum_{i, j = 1}^{d_\alpha} (\sigma_{ii}^2  - |\sigma_{ii}|^2) +\sum_{i\neq j = 1}^{d_\alpha}(\sigma_{ii} \sigma_{jj}- |\sigma_{ij}|^2 )\bigg \rangle_k.
\end{equation} 
Since $\sigma = \rho_k$ is Hermitian, its diagonal entries are real and $\sigma_{ii}^2 = |\sigma_{ii}|^2$. This gives us 
\begin{align}
    \langle f_2^{(d)}(\rho_{k,\alpha}) \rangle_k &= \frac{1}{2}\sum_{i\neq j = 1}^{d_\alpha}  \bigg( \langle \sigma_{ii} \sigma_{jj} \rangle_k-\langle |\sigma_{ij}|^2  \rangle_k\bigg) \nonumber \\
    &= \frac{d_\alpha(d_\alpha -1)}{2}\bigg( \langle \sigma_{ii} \sigma_{jj} \rangle_k-\langle |\sigma_{ij}|^2  \rangle_k\bigg),
\end{align}
since there are $d_\alpha (d_\alpha-1)$ order pairs $(i,j)$ and all terms in the sum are equal. The traces on the right can be found using the second- and fourth-moment identities for Haar random distributions~\cite{zyczkowski2001}
\begin{equation}
    \langle q_a q_b^* \rangle_k = \frac{\delta_{ab}}{N}, \qquad \langle q_a q_b q_c^* q_d^*  \rangle_k = \frac{\delta_{ac} \delta_{bd} + \delta_{ad} \delta_{bc}}{N(N+1)}.
\end{equation}
In particular,
\begin{align}
   \langle \sigma_{ii} \sigma_{jj} \rangle_k &= \frac{N}{d^2(N+1)}, \\
   \langle |\sigma_{ij}|^2  \rangle_k &= \frac{1}{d(N+1)}.
\end{align}
Finally, we can substitute these into our expression for $\langle f_2^{(d)}(\rho_{k,\alpha}) \rangle_k $, on on simplifying and substituting $n$ = $d^n$, we get
\begin{equation}
    \langle f_2^{(d)}(\rho_{k,\alpha}) \rangle_k = \frac{d_\alpha(d_\alpha-1)}{2} \frac{d^{n-1}-1}{d(d^n + 1)}.
\end{equation}
This yields
\begin{equation}
    Q_\alpha = \frac{2d}{d-1}\langle f_2^{(d)}(\rho_{k,\alpha}) \rangle_k = \frac{d_\alpha(d_\alpha-1)}{d(d-1)} \bigg(\frac{1-d^{1-n}}{1 + d^{-n}} \bigg),
\end{equation}
as quoted in Eq.~\eqref{eq:equipartition_intra}.

For the interference contributions, a similar technique can be used to probe $\langle f_2^{(d)}(\rho_{k,\alpha\beta}) \rangle_k $. Since $\Tr \rho_{k,\alpha\beta} = 0$, and $\Tr \rho_{k,\alpha\beta}^2 = 2\Tr (\rho_k P_\alpha \rho_k P_\beta)$, we have
\begin{equation}
\label{eq:f2inter}
    \langle f_2^{(d)}(\rho_{k,\alpha\beta}) \rangle_k = \frac{1}{2}\big(-2 \langle \Tr (\rho_k P_\alpha \rho_k P_\beta)\rangle_k\big) = - \langle \sum_{i=1}^{d_\alpha} \sum_{j=1}^{d_\beta} |\rho_{ij}|^2\rangle_k
\end{equation}
Once again, since all terms of the sum are equal and there are $d_\alpha d_\beta$ of such terms in total, we see
\begin{equation}
    \langle f_2^{(d)}(\rho_{k,\alpha\beta}) \rangle_k = -\frac{d_\alpha d_\beta}{d(d^n + 1)}.
\end{equation}
Next, note that 
\begin{equation}
    \langle p_{k,\alpha} p_{k,\beta} \rangle_k = \langle \Tr \rho_{k,\alpha} \Tr \rho_{k,\beta}\rangle_k = \sum_{i=1}^{d_\alpha} \sum_{j=1}^{d_\beta}\langle \sigma_{ii} \sigma_{jj}\rangle_k = d_\alpha d_\beta \frac{N}{d^2(N+1)} = \frac{d_\alpha d_\beta}{d^2} \frac{1}{1+d^{-n}}
\end{equation}

Finally, using Eq.~\eqref{eq:final_inter} and substituting both the above averages, we get
\begin{align}
    Q_{\alpha\beta} &= \frac{2d}{d-1} \bigg( \frac{d_\alpha d_\beta}{d^2}\frac{d^n}{d^{n}+1} -\frac{d_\alpha d_\beta}{d(d^n + 1)} \bigg) \nonumber \\
    &= \frac{2 d_\alpha d_\beta}{d(d-1)} \bigg( \frac{d^n}{d^n + 1} - \frac{d}{d^n + 1}\bigg) \nonumber \\
    & = \frac{2 d_\alpha d_\beta}{d(d-1)} \bigg( \frac{1 - d^{1-n}}{1+ d^{-n}}\bigg).
\end{align}
This is precisely Eq.~\eqref{eq:equipartition_inter} in the main text, and we see that the finite size correction is exactly the same as that for the sector-wise contributions $Q_\alpha$
   
\section{Experimental protocol to measure interference contributions}
\label{app:c}
In this section we describe a possible method to verify Eq.~\eqref{eq:final_inter}. Given a state $\ket{\psi}$, we can act on each particle by $(P_\mu + P_\nu)$ in order to destroy all contributions but those from sectors $\mu, \nu$ and interference between them. To see this, note that
\begin{equation}
    \rho_k \rightarrow (P_\mu + P_\nu) \rho_k (P_\mu + P_\nu) = \rho_{k,\mu} + \rho_{k,\nu} + \rho_{k,\mu\nu}.
\end{equation}
This further implies that all of $p_{k,\alpha}, \Tr(\rho_{k,\alpha}^2)$ and $\Tr(\rho_{k,\alpha\beta}^2)$ remain unchanged if and only if $\alpha = \mu$ and $\beta = \nu$. All other values of these quantities vanish, leaving (compare Eq.~\eqref{eq:Q_exp})
\begin{align}
    Q' &= \frac{d}{d-1}\Big\langle  \bigg( p_{k,\mu}^2 - \Tr \rho_{k,\mu}^2 \bigg) \Big\rangle_k + \frac{d}{d-1}\Big\langle  \bigg( p_{k,\nu}^2 - \Tr \rho_{k,\nu}^2 \bigg)\Big\rangle_k + \frac{d}{d-1} \Big\langle \bigg( 2p_{k,\mu}p_{k,\nu} - \Tr \rho_{k,\mu \nu}^2 \bigg)\Big\rangle_k \nonumber \\
    &= Q_\mu + Q_\nu + Q_{\mu\nu}.
\end{align}
This allows $Q_{\mu\nu}$ to be recovered experimentally after measuring all of $Q', Q_\mu$ and $Q_\nu$ using previously described methods. On the other hand, in order to evaluate the RHS, we need to evaluate $p_{k,\alpha}$ and $f_2^{(d)}(\rho_{k,\alpha\beta}) = -\Tr(\rho_k P_\alpha\rho_k P_\beta)$, see Eq.~\eqref{eq:f2inter}. While the probabilities $p_\alpha$ can directly be evaluated by repeated charge measurement on copies of the state, to evaluate $\Tr(\rho_k P_\alpha\rho_k P_\beta)$ we can use the quantum circuit shown in Fig.~\ref{fig:circuit}. 

An ancillary register of $n$ qubits prepared in the state $\ket{0}^{\otimes n}$ is treated with Hadamard gates, and two copies of the system state $\ket{\psi}$ are subjected to controlled-$U$ operations. In particular, the $k^\mathrm{th}$ ancilla is used as a control for the operator 
\begin{equation}
    U_{k,\alpha\beta} = (P_\alpha \otimes P_\beta)\swap,
\end{equation}
which acts on the $k^\mathrm{th}$ particle of the system as its target. Here $\swap$ is a two-particle swap operator, which acts as 
\begin{equation}
    \swap (\rho_A \otimes \rho_B) = \rho_B \otimes \rho_A.
\end{equation}
Together, the $k^\mathrm{th}$ ancilla and the $k^\mathrm{th}$ particle evolve under the circuit as
\begin{align}
    \ket{0}\bra{0} \otimes \rho_k \otimes \rho_k \xrightarrow{H\otimes \mathbbm{1} \otimes \mathbbm{1}} & \ket{+}\bra{+} \otimes \rho_k \otimes \rho_k \nonumber \\
    \xrightarrow{\mathrm{C-}U_{k,\alpha\beta}} & \;\frac{1}{2} \bigg(\ket{0}\bra{0} \otimes \rho_k \otimes \rho_k \nonumber \\ & 
    + \ket{0}\bra{1} \otimes (\rho_k \otimes \rho_k)U^\dagger_{k,\alpha\beta} \nonumber \\ & 
    + \ket{1}\bra{0} \otimes U_{k,\alpha\beta}(\rho_k \otimes \rho_k) \nonumber \\& 
    + \ket{1}\bra{1} \otimes U_{k,\alpha\beta}(\rho_k \otimes \rho_k)U^\dagger_{k,\alpha\beta}\bigg) \nonumber \\
    \xrightarrow{H \otimes \mathbbm{1} \otimes \mathbbm{1}} & \;\frac{1}{4} \bigg( \ket{0}\bra{0} \otimes \big((1 + U_{k,\alpha\beta}) (\rho_k \otimes \rho_k) (1 + U_{k,\alpha\beta})\big) \nonumber \\ &
    + \ket{0}\bra{1} \otimes\big((1  + U^\dagger_{k,\alpha\beta}) (\rho_k \otimes \rho_k) (1 + U_{k,\alpha\beta})\big) \nonumber \\ &
    + \ket{1}\bra{0} \otimes \big((1 - U_{k,\alpha\beta}) (\rho_k \otimes \rho_k) (1 + U^\dagger_{k,\alpha\beta})\big) \nonumber \\ &
    + \ket{1}\bra{1} \otimes \big((1 - U_{k,\alpha\beta}) (\rho_k \otimes \rho_k) (1 - U^\dagger_{k,\alpha\beta})\big)\bigg). \nonumber 
\end{align}

Now, to find the expectation value of the $Z$ measurement, we first need to trace out the non-ancillary registers. We get the RDM of the ancilla to be
\begin{equation}
    \rho_{\mathrm{ancilla}} = \frac{1}{2} \bigg( \ket{0}\bra{0} \Tr\big(( \rho_k \otimes \rho_k) (1 + U_{k,\alpha\beta})\big) + \ket{1}\bra{1} \Tr\big(( \rho_k \otimes \rho_k) (1 - U_{k,\alpha\beta})\big) \bigg),
\end{equation}
where we have used the hermiticity of $U_{k,\alpha\beta}$ and the cyclicity of the trace, which causes the coefficients of $\ket{0}\bra{1}$ and $\ket{1}\bra{0}$ to vanish. Finally, the expectation value of the $Z$ measurement is given by 
\begin{align}
    \langle Z \rangle &= \Tr(\rho Z) \nonumber \\ 
    &= \Tr\big((\rho_k \otimes \rho_k) U_{k,\alpha\beta}\big) \nonumber \\
    &= \Tr \big((\rho_k \otimes \rho_k)(P_\alpha \otimes P_\beta) \swap)\big) \nonumber \\
    &= \Tr \big((\rho_kP_\alpha \otimes \rho_kP_\beta) \swap)\big) \nonumber \\
    &= \Tr(\rho_kP_\alpha \rho_kP_\beta),
\end{align}
where we used the identity $\Tr((A\otimes B)\swap) = \Tr(AB)$. We see that the result of the measurement on the $k^\mathrm{th}$ ancilla is precisely (the negative of) $f_2^{(d)}(\rho_{k,\alpha\beta})$. Obtaining results from all $n$ ancillary qubits allows us to reconstruct the entire RHS of Eq.~\eqref{eq:final_inter} when combined with measurements of $p_{k,\alpha}$. 
\end{appendix}

\bibliographystyle{unsrt}

\begin{thebibliography}{99}%
\makeatletter
\providecommand \@ifxundefined [1]{%
 \@ifx{#1\undefined}
}%
\providecommand \@ifnum [1]{%
 \ifnum #1\expandafter \@firstoftwo
 \else \expandafter \@secondoftwo
 \fi
}%
\providecommand \@ifx [1]{%
 \ifx #1\expandafter \@firstoftwo
 \else \expandafter \@secondoftwo
 \fi
}%
\providecommand \natexlab [1]{#1}%
\providecommand \enquote  [1]{``#1''}%
\providecommand \bibnamefont  [1]{#1}%
\providecommand \bibfnamefont [1]{#1}%
\providecommand \citenamefont [1]{#1}%
\providecommand \href@noop [0]{\@secondoftwo}%
\providecommand \href [0]{\begingroup \@sanitize@url \@href}%
\providecommand \@href[1]{\@@startlink{#1}\@@href}%
\providecommand \@@href[1]{\endgroup#1\@@endlink}%
\providecommand \@sanitize@url [0]{\catcode `\\12\catcode `\$12\catcode `\&12\catcode `\#12\catcode `\^12\catcode `\_12\catcode `\%12\relax}%
\providecommand \@@startlink[1]{}%
\providecommand \@@endlink[0]{}%
\providecommand \url  [0]{\begingroup\@sanitize@url \@url }%
\providecommand \@url [1]{\endgroup\@href {#1}{\urlprefix }}%
\providecommand \urlprefix  [0]{URL }%
\providecommand \Eprint [0]{\href }%
\providecommand \doibase [0]{https://doi.org/}%
\providecommand \selectlanguage [0]{\@gobble}%
\providecommand \bibinfo  [0]{\@secondoftwo}%
\providecommand \bibfield  [0]{\@secondoftwo}%
\providecommand \translation [1]{[#1]}%
\providecommand \BibitemOpen [0]{}%
\providecommand \bibitemStop [0]{}%
\providecommand \bibitemNoStop [0]{.\EOS\space}%
\providecommand \EOS [0]{\spacefactor3000\relax}%
\providecommand \BibitemShut  [1]{\csname bibitem#1\endcsname}%
\let\auto@bib@innerbib\@empty
\bibitem [{\citenamefont {Einstein}\ \emph {et~al.}(1935)\citenamefont {Einstein}, \citenamefont {Podolsky},\ and\ \citenamefont {Rosen}}]{einstein1935}%
  \BibitemOpen
  \bibfield  {author} {\bibinfo {author} {\bibfnamefont {A.}~\bibnamefont {Einstein}}, \bibinfo {author} {\bibfnamefont {B.}~\bibnamefont {Podolsky}},\ and\ \bibinfo {author} {\bibfnamefont {N.}~\bibnamefont {Rosen}},\ }\bibfield  {title} {\bibinfo {title} {{Can Quantum-Mechanical Description of Physical Reality Be Considered Complete?}},\ }\href {https://doi.org/10.1103/PhysRev.47.777} {\bibfield  {journal} {\bibinfo  {journal} {Phys. Rev.}\ }\textbf {\bibinfo {volume} {47}},\ \bibinfo {pages} {777} (\bibinfo {year} {1935})}\BibitemShut {NoStop}%
\bibitem [{\citenamefont {Schr{\"o}dinger}(1935)}]{schrodinger1935}%
  \BibitemOpen
  \bibfield  {author} {\bibinfo {author} {\bibfnamefont {E.}~\bibnamefont {Schr{\"o}dinger}},\ }\bibfield  {title} {\bibinfo {title} {{Discussion of probability relations between separated systems}},\ }in\ \href {https://doi.org/10.1017/9781139028493.014} {\emph {\bibinfo {booktitle} {{Mathematical Proceedings of the Cambridge Philosophical Society}}}},\ Vol.~\bibinfo {volume} {31}\ (\bibinfo {organization} {Cambridge University Press},\ \bibinfo {year} {1935})\ pp.\ \bibinfo {pages} {555--563}\BibitemShut {NoStop}%
\bibitem [{\citenamefont {Horodecki}\ \emph {et~al.}(2009)\citenamefont {Horodecki}, \citenamefont {Horodecki}, \citenamefont {Horodecki},\ and\ \citenamefont {Horodecki}}]{horodecki2009}%
  \BibitemOpen
  \bibfield  {author} {\bibinfo {author} {\bibfnamefont {R.}~\bibnamefont {Horodecki}}, \bibinfo {author} {\bibfnamefont {P.}~\bibnamefont {Horodecki}}, \bibinfo {author} {\bibfnamefont {M.}~\bibnamefont {Horodecki}},\ and\ \bibinfo {author} {\bibfnamefont {K.}~\bibnamefont {Horodecki}},\ }\bibfield  {title} {\bibinfo {title} {{Quantum entanglement}},\ }\href {https://doi.org/10.1103/RevModPhys.81.865} {\bibfield  {journal} {\bibinfo  {journal} {Rev. Mod. Phys.}\ }\textbf {\bibinfo {volume} {81}},\ \bibinfo {pages} {865} (\bibinfo {year} {2009})}\BibitemShut {NoStop}%
\bibitem [{\citenamefont {Amico}\ \emph {et~al.}(2008)\citenamefont {Amico}, \citenamefont {Fazio}, \citenamefont {Osterloh},\ and\ \citenamefont {Vedral}}]{amico2008}%
  \BibitemOpen
  \bibfield  {author} {\bibinfo {author} {\bibfnamefont {L.}~\bibnamefont {Amico}}, \bibinfo {author} {\bibfnamefont {R.}~\bibnamefont {Fazio}}, \bibinfo {author} {\bibfnamefont {A.}~\bibnamefont {Osterloh}},\ and\ \bibinfo {author} {\bibfnamefont {V.}~\bibnamefont {Vedral}},\ }\bibfield  {title} {\bibinfo {title} {{Entanglement in many-body systems}},\ }\href {https://doi.org/10.1103/RevModPhys.80.517} {\bibfield  {journal} {\bibinfo  {journal} {Rev. Mod. Phys.}\ }\textbf {\bibinfo {volume} {80}},\ \bibinfo {pages} {517} (\bibinfo {year} {2008})}\BibitemShut {NoStop}%
\bibitem [{\citenamefont {Bennett}\ \emph {et~al.}(1996)\citenamefont {Bennett}, \citenamefont {Brassard}, \citenamefont {Popescu}, \citenamefont {Schumacher}, \citenamefont {Smolin},\ and\ \citenamefont {Wootters}}]{bennett1996a}%
  \BibitemOpen
  \bibfield  {author} {\bibinfo {author} {\bibfnamefont {C.~H.}\ \bibnamefont {Bennett}}, \bibinfo {author} {\bibfnamefont {G.}~\bibnamefont {Brassard}}, \bibinfo {author} {\bibfnamefont {S.}~\bibnamefont {Popescu}}, \bibinfo {author} {\bibfnamefont {B.}~\bibnamefont {Schumacher}}, \bibinfo {author} {\bibfnamefont {J.~A.}\ \bibnamefont {Smolin}},\ and\ \bibinfo {author} {\bibfnamefont {W.~K.}\ \bibnamefont {Wootters}},\ }\bibfield  {title} {\bibinfo {title} {{Purification of Noisy Entanglement and Faithful Teleportation via Noisy Channels}},\ }\href {https://doi.org/10.1103/PhysRevLett.76.722} {\bibfield  {journal} {\bibinfo  {journal} {Phys. Rev. Lett.}\ }\textbf {\bibinfo {volume} {76}},\ \bibinfo {pages} {722} (\bibinfo {year} {1996})}\BibitemShut {NoStop}%
\bibitem [{\citenamefont {Cory}\ \emph {et~al.}(1998)\citenamefont {Cory}, \citenamefont {Price}, \citenamefont {Maas}, \citenamefont {Knill}, \citenamefont {Laflamme}, \citenamefont {Zurek}, \citenamefont {Havel},\ and\ \citenamefont {Somaroo}}]{cory1998}%
  \BibitemOpen
  \bibfield  {author} {\bibinfo {author} {\bibfnamefont {D.~G.}\ \bibnamefont {Cory}}, \bibinfo {author} {\bibfnamefont {M.~D.}\ \bibnamefont {Price}}, \bibinfo {author} {\bibfnamefont {W.}~\bibnamefont {Maas}}, \bibinfo {author} {\bibfnamefont {E.}~\bibnamefont {Knill}}, \bibinfo {author} {\bibfnamefont {R.}~\bibnamefont {Laflamme}}, \bibinfo {author} {\bibfnamefont {W.~H.}\ \bibnamefont {Zurek}}, \bibinfo {author} {\bibfnamefont {T.~F.}\ \bibnamefont {Havel}},\ and\ \bibinfo {author} {\bibfnamefont {S.~S.}\ \bibnamefont {Somaroo}},\ }\bibfield  {title} {\bibinfo {title} {{Experimental quantum error correction}},\ }\href {https://doi.org/10.1103/PhysRevLett.81.2152} {\bibfield  {journal} {\bibinfo  {journal} {Physical Review Letters}\ }\textbf {\bibinfo {volume} {81}},\ \bibinfo {pages} {2152} (\bibinfo {year} {1998})}\BibitemShut {NoStop}%
\bibitem [{\citenamefont {Schindler}\ \emph {et~al.}(2011)\citenamefont {Schindler}, \citenamefont {Barreiro}, \citenamefont {Monz}, \citenamefont {Nebendahl}, \citenamefont {Nigg}, \citenamefont {Chwalla}, \citenamefont {Hennrich},\ and\ \citenamefont {Blatt}}]{schindler2011}%
  \BibitemOpen
  \bibfield  {author} {\bibinfo {author} {\bibfnamefont {P.}~\bibnamefont {Schindler}}, \bibinfo {author} {\bibfnamefont {J.~T.}\ \bibnamefont {Barreiro}}, \bibinfo {author} {\bibfnamefont {T.}~\bibnamefont {Monz}}, \bibinfo {author} {\bibfnamefont {V.}~\bibnamefont {Nebendahl}}, \bibinfo {author} {\bibfnamefont {D.}~\bibnamefont {Nigg}}, \bibinfo {author} {\bibfnamefont {M.}~\bibnamefont {Chwalla}}, \bibinfo {author} {\bibfnamefont {M.}~\bibnamefont {Hennrich}},\ and\ \bibinfo {author} {\bibfnamefont {R.}~\bibnamefont {Blatt}},\ }\bibfield  {title} {\bibinfo {title} {{Experimental repetitive quantum error correction}},\ }\href {https://doi.org/10.1126/science.1203329} {\bibfield  {journal} {\bibinfo  {journal} {Science}\ }\textbf {\bibinfo {volume} {332}},\ \bibinfo {pages} {1059} (\bibinfo {year} {2011})}\BibitemShut {NoStop}%
\bibitem [{\citenamefont {and}(1994)}]{josza1994}%
  \BibitemOpen
  \bibfield  {author} {\bibinfo {author} {\bibfnamefont {R.~J.}\ \bibnamefont {and}},\ }\bibfield  {title} {\bibinfo {title} {{Fidelity for Mixed Quantum States}},\ }\href {https://doi.org/10.1080/09500349414552171} {\bibfield  {journal} {\bibinfo  {journal} {Journal of Modern Optics}\ }\textbf {\bibinfo {volume} {41}},\ \bibinfo {pages} {2315} (\bibinfo {year} {1994})},\ \Eprint {https://arxiv.org/abs/https://doi.org/10.1080/09500349414552171} {https://doi.org/10.1080/09500349414552171} \BibitemShut {NoStop}%
\bibitem [{\citenamefont {Gross}\ \emph {et~al.}(2009)\citenamefont {Gross}, \citenamefont {Flammia},\ and\ \citenamefont {Eisert}}]{gross2009}%
  \BibitemOpen
  \bibfield  {author} {\bibinfo {author} {\bibfnamefont {D.}~\bibnamefont {Gross}}, \bibinfo {author} {\bibfnamefont {S.~T.}\ \bibnamefont {Flammia}},\ and\ \bibinfo {author} {\bibfnamefont {J.}~\bibnamefont {Eisert}},\ }\bibfield  {title} {\bibinfo {title} {{Most Quantum States Are Too Entangled To Be Useful As Computational Resources}},\ }\href {https://doi.org/10.1103/PhysRevLett.102.190501} {\bibfield  {journal} {\bibinfo  {journal} {Phys. Rev. Lett.}\ }\textbf {\bibinfo {volume} {102}},\ \bibinfo {pages} {190501} (\bibinfo {year} {2009})}\BibitemShut {NoStop}%
\bibitem [{\citenamefont {Nielsen}\ and\ \citenamefont {Chuang}(2010)}]{nielsen2010}%
  \BibitemOpen
  \bibfield  {author} {\bibinfo {author} {\bibfnamefont {M.~A.}\ \bibnamefont {Nielsen}}\ and\ \bibinfo {author} {\bibfnamefont {I.~L.}\ \bibnamefont {Chuang}},\ }\href {https://doi.org/10.1017/CBO9780511976667} {\emph {\bibinfo {title} {{Quantum Computation and Quantum Information: 10th Anniversary Edition}}}}\ (\bibinfo  {publisher} {Cambridge University Press},\ \bibinfo {year} {2010})\BibitemShut {NoStop}%
\bibitem [{\citenamefont {Yamasaki}\ \emph {et~al.}(2018)\citenamefont {Yamasaki}, \citenamefont {Pirker}, \citenamefont {Murao}, \citenamefont {Dür},\ and\ \citenamefont {Kraus}}]{yamasaki2018}%
  \BibitemOpen
  \bibfield  {author} {\bibinfo {author} {\bibfnamefont {H.}~\bibnamefont {Yamasaki}}, \bibinfo {author} {\bibfnamefont {A.}~\bibnamefont {Pirker}}, \bibinfo {author} {\bibfnamefont {M.}~\bibnamefont {Murao}}, \bibinfo {author} {\bibfnamefont {W.}~\bibnamefont {Dür}},\ and\ \bibinfo {author} {\bibfnamefont {B.}~\bibnamefont {Kraus}},\ }\bibfield  {title} {\bibinfo {title} {{Multipartite entanglement outperforming bipartite entanglement under limited quantum system sizes}},\ }\bibfield  {journal} {\bibinfo  {journal} {Physical Review A}\ }\textbf {\bibinfo {volume} {98}},\ \href {https://doi.org/10.1103/physreva.98.052313} {10.1103/physreva.98.052313} (\bibinfo {year} {2018})\BibitemShut {NoStop}%
\bibitem [{\citenamefont {Boixo}\ \emph {et~al.}(2018)\citenamefont {Boixo}, \citenamefont {Isakov}, \citenamefont {Smelyanskiy}, \citenamefont {Babbush}, \citenamefont {Ding}, \citenamefont {Jiang}, \citenamefont {Bremner}, \citenamefont {Martinis},\ and\ \citenamefont {Neven}}]{boixo2018}%
  \BibitemOpen
  \bibfield  {author} {\bibinfo {author} {\bibfnamefont {S.}~\bibnamefont {Boixo}}, \bibinfo {author} {\bibfnamefont {S.~V.}\ \bibnamefont {Isakov}}, \bibinfo {author} {\bibfnamefont {V.~N.}\ \bibnamefont {Smelyanskiy}}, \bibinfo {author} {\bibfnamefont {R.}~\bibnamefont {Babbush}}, \bibinfo {author} {\bibfnamefont {N.}~\bibnamefont {Ding}}, \bibinfo {author} {\bibfnamefont {Z.}~\bibnamefont {Jiang}}, \bibinfo {author} {\bibfnamefont {M.~J.}\ \bibnamefont {Bremner}}, \bibinfo {author} {\bibfnamefont {J.~M.}\ \bibnamefont {Martinis}},\ and\ \bibinfo {author} {\bibfnamefont {H.}~\bibnamefont {Neven}},\ }\bibfield  {title} {\bibinfo {title} {{Characterizing quantum supremacy in near-term devices}},\ }\href {https://doi.org/10.1038/s41567-018-0124-x} {\bibfield  {journal} {\bibinfo  {journal} {Nature Physics}\ }\textbf {\bibinfo {volume} {14}},\ \bibinfo {pages} {595} (\bibinfo {year} {2018})}\BibitemShut {NoStop}%
\bibitem [{\citenamefont {Neill}\ \emph {et~al.}(2018)\citenamefont {Neill}, \citenamefont {Roushan}, \citenamefont {Kechedzhi}, \citenamefont {Boixo}, \citenamefont {Isakov}, \citenamefont {Smelyanskiy} \emph {et~al.}}]{neill2018}%
  \BibitemOpen
  \bibfield  {author} {\bibinfo {author} {\bibfnamefont {C.}~\bibnamefont {Neill}}, \bibinfo {author} {\bibfnamefont {P.}~\bibnamefont {Roushan}}, \bibinfo {author} {\bibfnamefont {K.}~\bibnamefont {Kechedzhi}}, \bibinfo {author} {\bibfnamefont {S.}~\bibnamefont {Boixo}}, \bibinfo {author} {\bibfnamefont {S.~V.}\ \bibnamefont {Isakov}}, \bibinfo {author} {\bibfnamefont {V.}~\bibnamefont {Smelyanskiy}}, \emph {et~al.},\ }\bibfield  {title} {\bibinfo {title} {{A blueprint for demonstrating quantum supremacy with superconducting qubits}},\ }\href {https://doi.org/10.1126/science.aao4309} {\bibfield  {journal} {\bibinfo  {journal} {Science}\ }\textbf {\bibinfo {volume} {360}},\ \bibinfo {pages} {195} (\bibinfo {year} {2018})}\BibitemShut {NoStop}%
\bibitem [{\citenamefont {Arute}\ \emph {et~al.}(2019)\citenamefont {Arute}, \citenamefont {Arya}, \citenamefont {Babbush}, \citenamefont {Bacon}, \citenamefont {Bardin}, \citenamefont {Barends} \emph {et~al.}}]{arute2019}%
  \BibitemOpen
  \bibfield  {author} {\bibinfo {author} {\bibfnamefont {F.}~\bibnamefont {Arute}}, \bibinfo {author} {\bibfnamefont {K.}~\bibnamefont {Arya}}, \bibinfo {author} {\bibfnamefont {R.}~\bibnamefont {Babbush}}, \bibinfo {author} {\bibfnamefont {D.}~\bibnamefont {Bacon}}, \bibinfo {author} {\bibfnamefont {J.~C.}\ \bibnamefont {Bardin}}, \bibinfo {author} {\bibfnamefont {R.}~\bibnamefont {Barends}}, \emph {et~al.},\ }\bibfield  {title} {\bibinfo {title} {{Quantum supremacy using a programmable superconducting processor}},\ }\href {https://doi.org/10.1038/s41586-019-1666-5} {\bibfield  {journal} {\bibinfo  {journal} {Nature}\ }\textbf {\bibinfo {volume} {574}},\ \bibinfo {pages} {505} (\bibinfo {year} {2019})}\BibitemShut {NoStop}%
\bibitem [{\citenamefont {Basko}\ \emph {et~al.}(2006)\citenamefont {Basko}, \citenamefont {Aleiner},\ and\ \citenamefont {Altshuler}}]{Basko2006}%
  \BibitemOpen
  \bibfield  {author} {\bibinfo {author} {\bibfnamefont {D.~M.}\ \bibnamefont {Basko}}, \bibinfo {author} {\bibfnamefont {I.~L.}\ \bibnamefont {Aleiner}},\ and\ \bibinfo {author} {\bibfnamefont {B.~L.}\ \bibnamefont {Altshuler}},\ }\bibfield  {title} {\bibinfo {title} {{Metal–insulator transition in a weakly interacting many-electron system with localized single-particle states}},\ }\href {https://doi.org/10.1016/j.aop.2005.11.014} {\bibfield  {journal} {\bibinfo  {journal} {Annals of Physics}\ }\textbf {\bibinfo {volume} {321}},\ \bibinfo {pages} {1126} (\bibinfo {year} {2006})}\BibitemShut {NoStop}%
\bibitem [{\citenamefont {Nandkishore}\ and\ \citenamefont {Huse}(2015)}]{Nandkishore2015}%
  \BibitemOpen
  \bibfield  {author} {\bibinfo {author} {\bibfnamefont {R.}~\bibnamefont {Nandkishore}}\ and\ \bibinfo {author} {\bibfnamefont {D.~A.}\ \bibnamefont {Huse}},\ }\bibfield  {title} {\bibinfo {title} {{Many-body localization and thermalization in quantum statistical mechanics}},\ }\href {https://doi.org/10.1146/annurev-conmatphys-031214-014726} {\bibfield  {journal} {\bibinfo  {journal} {Annual Review of Condensed Matter Physics}\ }\textbf {\bibinfo {volume} {6}},\ \bibinfo {pages} {15} (\bibinfo {year} {2015})}\BibitemShut {NoStop}%
\bibitem [{\citenamefont {Štrkalj}\ \emph {et~al.}(2021)\citenamefont {Štrkalj}, \citenamefont {Doggen}, \citenamefont {Gornyi},\ and\ \citenamefont {Zilberberg}}]{Strkalj2021}%
  \BibitemOpen
  \bibfield  {author} {\bibinfo {author} {\bibfnamefont {A.}~\bibnamefont {Štrkalj}}, \bibinfo {author} {\bibfnamefont {E.~V.~H.}\ \bibnamefont {Doggen}}, \bibinfo {author} {\bibfnamefont {I.~V.}\ \bibnamefont {Gornyi}},\ and\ \bibinfo {author} {\bibfnamefont {O.}~\bibnamefont {Zilberberg}},\ }\bibfield  {title} {\bibinfo {title} {{Many-body localization in the interpolating aubry-andré-fibonacci model}},\ }\href {https://doi.org/10.1103/PhysRevResearch.3.033257} {\bibfield  {journal} {\bibinfo  {journal} {Phys. Rev. Research}\ }\textbf {\bibinfo {volume} {3}},\ \bibinfo {pages} {033257} (\bibinfo {year} {2021})}\BibitemShut {NoStop}%
\bibitem [{\citenamefont {Ferguson}\ \emph {et~al.}(2020)\citenamefont {Ferguson}, \citenamefont {Camenzind}, \citenamefont {Müller}, \citenamefont {Biesinger}, \citenamefont {Scheller}, \citenamefont {Braunecker}, \citenamefont {Zumbühl},\ and\ \citenamefont {Zilberberg}}]{Ferguson2020}%
  \BibitemOpen
  \bibfield  {author} {\bibinfo {author} {\bibfnamefont {M.~S.}\ \bibnamefont {Ferguson}}, \bibinfo {author} {\bibfnamefont {L.~C.}\ \bibnamefont {Camenzind}}, \bibinfo {author} {\bibfnamefont {C.}~\bibnamefont {Müller}}, \bibinfo {author} {\bibfnamefont {D.~E.~F.}\ \bibnamefont {Biesinger}}, \bibinfo {author} {\bibfnamefont {C.~P.}\ \bibnamefont {Scheller}}, \bibinfo {author} {\bibfnamefont {B.}~\bibnamefont {Braunecker}}, \bibinfo {author} {\bibfnamefont {D.~M.}\ \bibnamefont {Zumbühl}},\ and\ \bibinfo {author} {\bibfnamefont {O.}~\bibnamefont {Zilberberg}},\ }\href {https://arxiv.org/abs/2010.04635} {\bibinfo {title} {{Quantum measurement induces a many-body transition}}} (\bibinfo {year} {2020}),\ \Eprint {https://arxiv.org/abs/2010.04635} {arXiv:2010.04635 [cond-mat.mes-hall]} \BibitemShut {NoStop}%
\bibitem [{\citenamefont {Li}\ \emph {et~al.}(2018)\citenamefont {Li}, \citenamefont {Chen},\ and\ \citenamefont {Fisher}}]{liy2018}%
  \BibitemOpen
  \bibfield  {author} {\bibinfo {author} {\bibfnamefont {Y.}~\bibnamefont {Li}}, \bibinfo {author} {\bibfnamefont {X.}~\bibnamefont {Chen}},\ and\ \bibinfo {author} {\bibfnamefont {M.~P.~A.}\ \bibnamefont {Fisher}},\ }\bibfield  {title} {\bibinfo {title} {{Quantum zeno effect and the many-body entanglement transition}},\ }\href {https://doi.org/10.1103/PhysRevB.98.205136} {\bibfield  {journal} {\bibinfo  {journal} {Phys. Rev. B}\ }\textbf {\bibinfo {volume} {98}},\ \bibinfo {pages} {205136} (\bibinfo {year} {2018})}\BibitemShut {NoStop}%
\bibitem [{\citenamefont {Szyniszewski}\ \emph {et~al.}(2019)\citenamefont {Szyniszewski}, \citenamefont {Romito},\ and\ \citenamefont {Schomerus}}]{Szyniszewski2019}%
  \BibitemOpen
  \bibfield  {author} {\bibinfo {author} {\bibfnamefont {M.}~\bibnamefont {Szyniszewski}}, \bibinfo {author} {\bibfnamefont {A.}~\bibnamefont {Romito}},\ and\ \bibinfo {author} {\bibfnamefont {H.}~\bibnamefont {Schomerus}},\ }\bibfield  {title} {\bibinfo {title} {{Entanglement transition from variable-strength weak measurements}},\ }\href {https://doi.org/10.1103/PhysRevB.100.064204} {\bibfield  {journal} {\bibinfo  {journal} {Phys. Rev. B}\ }\textbf {\bibinfo {volume} {100}},\ \bibinfo {pages} {064204} (\bibinfo {year} {2019})}\BibitemShut {NoStop}%
\bibitem [{\citenamefont {Shapourian}\ \emph {et~al.}(2017)\citenamefont {Shapourian}, \citenamefont {Shiozaki},\ and\ \citenamefont {Ryu}}]{Shapourian2017}%
  \BibitemOpen
  \bibfield  {author} {\bibinfo {author} {\bibfnamefont {H.}~\bibnamefont {Shapourian}}, \bibinfo {author} {\bibfnamefont {K.}~\bibnamefont {Shiozaki}},\ and\ \bibinfo {author} {\bibfnamefont {S.}~\bibnamefont {Ryu}},\ }\bibfield  {title} {\bibinfo {title} {{Partial time-reversal transformation and entanglement negativity in fermionic systems}},\ }\href {https://doi.org/10.1103/PhysRevB.95.165101} {\bibfield  {journal} {\bibinfo  {journal} {Phys. Rev. B}\ }\textbf {\bibinfo {volume} {95}},\ \bibinfo {pages} {165101} (\bibinfo {year} {2017})}\BibitemShut {NoStop}%
\bibitem [{\citenamefont {Campagnano}\ \emph {et~al.}(2012)\citenamefont {Campagnano}, \citenamefont {Zilberberg}, \citenamefont {Gornyi}, \citenamefont {Feldman}, \citenamefont {Potter},\ and\ \citenamefont {Gefen}}]{Campagnano2012}%
  \BibitemOpen
  \bibfield  {author} {\bibinfo {author} {\bibfnamefont {G.}~\bibnamefont {Campagnano}}, \bibinfo {author} {\bibfnamefont {O.}~\bibnamefont {Zilberberg}}, \bibinfo {author} {\bibfnamefont {I.~V.}\ \bibnamefont {Gornyi}}, \bibinfo {author} {\bibfnamefont {D.~E.}\ \bibnamefont {Feldman}}, \bibinfo {author} {\bibfnamefont {A.~C.}\ \bibnamefont {Potter}},\ and\ \bibinfo {author} {\bibfnamefont {Y.}~\bibnamefont {Gefen}},\ }\bibfield  {title} {\bibinfo {title} {{Hanbury brown–twiss interference of anyons}},\ }\href {https://doi.org/10.1103/PhysRevLett.109.106802} {\bibfield  {journal} {\bibinfo  {journal} {Phys. Rev. Lett.}\ }\textbf {\bibinfo {volume} {109}},\ \bibinfo {pages} {106802} (\bibinfo {year} {2012})}\BibitemShut {NoStop}%
\bibitem [{\citenamefont {Liu}\ \emph {et~al.}(2022)\citenamefont {Liu}, \citenamefont {Sohal}, \citenamefont {Kudler-Flam},\ and\ \citenamefont {Ryu}}]{Liu2022}%
  \BibitemOpen
  \bibfield  {author} {\bibinfo {author} {\bibfnamefont {Y.}~\bibnamefont {Liu}}, \bibinfo {author} {\bibfnamefont {R.}~\bibnamefont {Sohal}}, \bibinfo {author} {\bibfnamefont {J.}~\bibnamefont {Kudler-Flam}},\ and\ \bibinfo {author} {\bibfnamefont {S.}~\bibnamefont {Ryu}},\ }\bibfield  {title} {\bibinfo {title} {{Multipartitioning topological phases by vertex states and quantum entanglement}},\ }\href {https://doi.org/10.1103/PhysRevB.105.115107} {\bibfield  {journal} {\bibinfo  {journal} {Phys. Rev. B}\ }\textbf {\bibinfo {volume} {105}},\ \bibinfo {pages} {115107} (\bibinfo {year} {2022})}\BibitemShut {NoStop}%
\bibitem [{\citenamefont {Vidal}(2000)}]{vidal2000}%
  \BibitemOpen
  \bibfield  {author} {\bibinfo {author} {\bibfnamefont {G.}~\bibnamefont {Vidal}},\ }\bibfield  {title} {\bibinfo {title} {{Entanglement monotones}},\ }\href {https://doi.org/10.1080/09500340008244048} {\bibfield  {journal} {\bibinfo  {journal} {Journal of Modern Optics}\ }\textbf {\bibinfo {volume} {47}},\ \bibinfo {pages} {355–376} (\bibinfo {year} {2000})}\BibitemShut {NoStop}%
\bibitem [{\citenamefont {Walter}\ \emph {et~al.}(2013)\citenamefont {Walter}, \citenamefont {Doran}, \citenamefont {Gross},\ and\ \citenamefont {Christandl}}]{walter2013}%
  \BibitemOpen
  \bibfield  {author} {\bibinfo {author} {\bibfnamefont {M.}~\bibnamefont {Walter}}, \bibinfo {author} {\bibfnamefont {B.}~\bibnamefont {Doran}}, \bibinfo {author} {\bibfnamefont {D.}~\bibnamefont {Gross}},\ and\ \bibinfo {author} {\bibfnamefont {M.}~\bibnamefont {Christandl}},\ }\bibfield  {title} {\bibinfo {title} {{Entanglement Polytopes: Multiparticle Entanglement from Single-Particle Information}},\ }\href {https://doi.org/10.1126/science.1232957} {\bibfield  {journal} {\bibinfo  {journal} {Science}\ }\textbf {\bibinfo {volume} {340}},\ \bibinfo {pages} {1205} (\bibinfo {year} {2013})},\ \Eprint {https://arxiv.org/abs/https://www.science.org/doi/pdf/10.1126/science.1232957} {https://www.science.org/doi/pdf/10.1126/science.1232957} \BibitemShut {NoStop}%
\bibitem [{\citenamefont {Maci\k{a}\.{z}ek}\ and\ \citenamefont {Sawicki}(2018)}]{maciazek2018}%
  \BibitemOpen
  \bibfield  {author} {\bibinfo {author} {\bibfnamefont {T.}~\bibnamefont {Maci\k{a}\.{z}ek}}\ and\ \bibinfo {author} {\bibfnamefont {A.}~\bibnamefont {Sawicki}},\ }\bibfield  {title} {\bibinfo {title} {{Asymptotic properties of entanglement polytopes for large number of qubits}},\ }\href {https://doi.org/10.1088/1751-8121/aaa4d7} {\bibfield  {journal} {\bibinfo  {journal} {Journal of Physics A: Mathematical and Theoretical}\ }\textbf {\bibinfo {volume} {51}},\ \bibinfo {pages} {07LT01} (\bibinfo {year} {2018})}\BibitemShut {NoStop}%
\bibitem [{\citenamefont {Zhang}\ \emph {et~al.}(2018)\citenamefont {Zhang}, \citenamefont {Zeng}, \citenamefont {Fan}, \citenamefont {You},\ and\ \citenamefont {Nori}}]{zhang2018}%
  \BibitemOpen
  \bibfield  {author} {\bibinfo {author} {\bibfnamefont {Y.-R.}\ \bibnamefont {Zhang}}, \bibinfo {author} {\bibfnamefont {Y.}~\bibnamefont {Zeng}}, \bibinfo {author} {\bibfnamefont {H.}~\bibnamefont {Fan}}, \bibinfo {author} {\bibfnamefont {J.~Q.}\ \bibnamefont {You}},\ and\ \bibinfo {author} {\bibfnamefont {F.}~\bibnamefont {Nori}},\ }\bibfield  {title} {\bibinfo {title} {{Characterization of Topological States via Dual Multipartite Entanglement}},\ }\href {https://doi.org/10.1103/PhysRevLett.120.250501} {\bibfield  {journal} {\bibinfo  {journal} {Phys. Rev. Lett.}\ }\textbf {\bibinfo {volume} {120}},\ \bibinfo {pages} {250501} (\bibinfo {year} {2018})}\BibitemShut {NoStop}%
\bibitem [{\citenamefont {Pezz{\`e}}\ \emph {et~al.}(2017)\citenamefont {Pezz{\`e}}, \citenamefont {Gabbrielli}, \citenamefont {Lepori},\ and\ \citenamefont {Smerzi}}]{pezz2017}%
  \BibitemOpen
  \bibfield  {author} {\bibinfo {author} {\bibfnamefont {L.}~\bibnamefont {Pezz{\`e}}}, \bibinfo {author} {\bibfnamefont {M.}~\bibnamefont {Gabbrielli}}, \bibinfo {author} {\bibfnamefont {L.}~\bibnamefont {Lepori}},\ and\ \bibinfo {author} {\bibfnamefont {A.}~\bibnamefont {Smerzi}},\ }\bibfield  {title} {\bibinfo {title} {{Multipartite Entanglement in Topological Quantum Phases.}},\ }\href {https://api.semanticscholar.org/CorpusID:5482852} {\bibfield  {journal} {\bibinfo  {journal} {Physical review letters}\ }\textbf {\bibinfo {volume} {119 25}},\ \bibinfo {pages} {250401} (\bibinfo {year} {2017})}\BibitemShut {NoStop}%
\bibitem [{\citenamefont {Azses}\ \emph {et~al.}(2021)\citenamefont {Azses}, \citenamefont {Dalla~Torre},\ and\ \citenamefont {Sela}}]{azses2021}%
  \BibitemOpen
  \bibfield  {author} {\bibinfo {author} {\bibfnamefont {D.}~\bibnamefont {Azses}}, \bibinfo {author} {\bibfnamefont {E.~G.}\ \bibnamefont {Dalla~Torre}},\ and\ \bibinfo {author} {\bibfnamefont {E.}~\bibnamefont {Sela}},\ }\bibfield  {title} {\bibinfo {title} {{Observing Floquet topological order by symmetry resolution}},\ }\href {https://doi.org/10.1103/PhysRevB.104.L220301} {\bibfield  {journal} {\bibinfo  {journal} {Phys. Rev. B}\ }\textbf {\bibinfo {volume} {104}},\ \bibinfo {pages} {L220301} (\bibinfo {year} {2021})}\BibitemShut {NoStop}%
\bibitem [{\citenamefont {Pirmoradian}\ and\ \citenamefont {Tanhayi}(2024)}]{pirmoradian2024}%
  \BibitemOpen
  \bibfield  {author} {\bibinfo {author} {\bibfnamefont {R.}~\bibnamefont {Pirmoradian}}\ and\ \bibinfo {author} {\bibfnamefont {M.~R.}\ \bibnamefont {Tanhayi}},\ }\bibfield  {title} {\bibinfo {title} {{Symmetry-resolved entanglement entropy for local and non-local QFTs}},\ }\href {https://doi.org/10.1140/epjc/s10052-024-13212-8} {\bibfield  {journal} {\bibinfo  {journal} {The European Physical Journal C}\ }\textbf {\bibinfo {volume} {84}},\ \bibinfo {pages} {849} (\bibinfo {year} {2024})}\BibitemShut {NoStop}%
\bibitem [{\citenamefont {Vitale}\ \emph {et~al.}(2022)\citenamefont {Vitale}, \citenamefont {Elben}, \citenamefont {Kueng}, \citenamefont {Neven}, \citenamefont {Carrasco}, \citenamefont {Kraus}, \citenamefont {Zoller}, \citenamefont {Calabrese}, \citenamefont {Vermersch},\ and\ \citenamefont {Dalmonte}}]{vitale2022}%
  \BibitemOpen
  \bibfield  {author} {\bibinfo {author} {\bibfnamefont {V.}~\bibnamefont {Vitale}}, \bibinfo {author} {\bibfnamefont {A.}~\bibnamefont {Elben}}, \bibinfo {author} {\bibfnamefont {R.}~\bibnamefont {Kueng}}, \bibinfo {author} {\bibfnamefont {A.}~\bibnamefont {Neven}}, \bibinfo {author} {\bibfnamefont {J.}~\bibnamefont {Carrasco}}, \bibinfo {author} {\bibfnamefont {B.}~\bibnamefont {Kraus}}, \bibinfo {author} {\bibfnamefont {P.}~\bibnamefont {Zoller}}, \bibinfo {author} {\bibfnamefont {P.}~\bibnamefont {Calabrese}}, \bibinfo {author} {\bibfnamefont {B.}~\bibnamefont {Vermersch}},\ and\ \bibinfo {author} {\bibfnamefont {M.}~\bibnamefont {Dalmonte}},\ }\bibfield  {title} {\bibinfo {title} {{Symmetry-resolved dynamical purification in synthetic quantum matter}},\ }\href {https://doi.org/10.21468/SciPostPhys.12.3.106} {\bibfield  {journal} {\bibinfo  {journal} {SciPost Phys.}\ }\textbf {\bibinfo {volume} {12}},\ \bibinfo {pages} {106} (\bibinfo {year} {2022})}\BibitemShut {NoStop}%
\bibitem [{\citenamefont {Rath}\ \emph {et~al.}(2023)\citenamefont {Rath}, \citenamefont {Vitale}, \citenamefont {Murciano}, \citenamefont {Votto}, \citenamefont {Dubail}, \citenamefont {Kueng}, \citenamefont {Branciard}, \citenamefont {Calabrese},\ and\ \citenamefont {Vermersch}}]{rath2023}%
  \BibitemOpen
  \bibfield  {author} {\bibinfo {author} {\bibfnamefont {A.}~\bibnamefont {Rath}}, \bibinfo {author} {\bibfnamefont {V.}~\bibnamefont {Vitale}}, \bibinfo {author} {\bibfnamefont {S.}~\bibnamefont {Murciano}}, \bibinfo {author} {\bibfnamefont {M.}~\bibnamefont {Votto}}, \bibinfo {author} {\bibfnamefont {J.}~\bibnamefont {Dubail}}, \bibinfo {author} {\bibfnamefont {R.}~\bibnamefont {Kueng}}, \bibinfo {author} {\bibfnamefont {C.}~\bibnamefont {Branciard}}, \bibinfo {author} {\bibfnamefont {P.}~\bibnamefont {Calabrese}},\ and\ \bibinfo {author} {\bibfnamefont {B.}~\bibnamefont {Vermersch}},\ }\bibfield  {title} {\bibinfo {title} {{Entanglement Barrier and its Symmetry Resolution: Theory and Experimental Observation}},\ }\href {https://doi.org/10.1103/PRXQuantum.4.010318} {\bibfield  {journal} {\bibinfo  {journal} {PRX Quantum}\ }\textbf {\bibinfo {volume} {4}},\ \bibinfo {pages} {010318} (\bibinfo {year} {2023})}\BibitemShut {NoStop}%
\bibitem [{\citenamefont {Fossati}\ \emph {et~al.}(2023)\citenamefont {Fossati}, \citenamefont {Ares},\ and\ \citenamefont {Calabrese}}]{fossati2023}%
  \BibitemOpen
  \bibfield  {author} {\bibinfo {author} {\bibfnamefont {M.}~\bibnamefont {Fossati}}, \bibinfo {author} {\bibfnamefont {F.}~\bibnamefont {Ares}},\ and\ \bibinfo {author} {\bibfnamefont {P.}~\bibnamefont {Calabrese}},\ }\bibfield  {title} {\bibinfo {title} {{Symmetry-resolved entanglement in critical non-Hermitian systems}},\ }\href {https://doi.org/10.1103/PhysRevB.107.205153} {\bibfield  {journal} {\bibinfo  {journal} {Phys. Rev. B}\ }\textbf {\bibinfo {volume} {107}},\ \bibinfo {pages} {205153} (\bibinfo {year} {2023})}\BibitemShut {NoStop}%
\bibitem [{\citenamefont {Azses}\ and\ \citenamefont {Sela}(2020)}]{azses2020a}%
  \BibitemOpen
  \bibfield  {author} {\bibinfo {author} {\bibfnamefont {D.}~\bibnamefont {Azses}}\ and\ \bibinfo {author} {\bibfnamefont {E.}~\bibnamefont {Sela}},\ }\bibfield  {title} {\bibinfo {title} {{Symmetry-resolved entanglement in symmetry-protected topological phases}},\ }\bibfield  {journal} {\bibinfo  {journal} {Physical Review B}\ }\textbf {\bibinfo {volume} {102}},\ \href {https://doi.org/10.1103/physrevb.102.235157} {10.1103/physrevb.102.235157} (\bibinfo {year} {2020})\BibitemShut {NoStop}%
\bibitem [{\citenamefont {Azses}\ \emph {et~al.}(2020)\citenamefont {Azses}, \citenamefont {Haenel}, \citenamefont {Naveh}, \citenamefont {Raussendorf}, \citenamefont {Sela},\ and\ \citenamefont {Dalla~Torre}}]{azses2020b}%
  \BibitemOpen
  \bibfield  {author} {\bibinfo {author} {\bibfnamefont {D.}~\bibnamefont {Azses}}, \bibinfo {author} {\bibfnamefont {R.}~\bibnamefont {Haenel}}, \bibinfo {author} {\bibfnamefont {Y.}~\bibnamefont {Naveh}}, \bibinfo {author} {\bibfnamefont {R.}~\bibnamefont {Raussendorf}}, \bibinfo {author} {\bibfnamefont {E.}~\bibnamefont {Sela}},\ and\ \bibinfo {author} {\bibfnamefont {E.~G.}\ \bibnamefont {Dalla~Torre}},\ }\bibfield  {title} {\bibinfo {title} {{Identification of Symmetry-Protected Topological States on Noisy Quantum Computers}},\ }\bibfield  {journal} {\bibinfo  {journal} {Physical Review Letters}\ }\textbf {\bibinfo {volume} {125}},\ \href {https://doi.org/10.1103/physrevlett.125.120502} {10.1103/physrevlett.125.120502} (\bibinfo {year} {2020})\BibitemShut {NoStop}%
\bibitem [{\citenamefont {Azses}\ \emph {et~al.}(2023)\citenamefont {Azses}, \citenamefont {Mross},\ and\ \citenamefont {Sela}}]{azses2024}%
  \BibitemOpen
  \bibfield  {author} {\bibinfo {author} {\bibfnamefont {D.}~\bibnamefont {Azses}}, \bibinfo {author} {\bibfnamefont {D.~F.}\ \bibnamefont {Mross}},\ and\ \bibinfo {author} {\bibfnamefont {E.}~\bibnamefont {Sela}},\ }\bibfield  {title} {\bibinfo {title} {{Symmetry-resolved entanglement of two-dimensional symmetry-protected topological states}},\ }\href {https://doi.org/10.1103/PhysRevB.107.115113} {\bibfield  {journal} {\bibinfo  {journal} {Phys. Rev. B}\ }\textbf {\bibinfo {volume} {107}},\ \bibinfo {pages} {115113} (\bibinfo {year} {2023})}\BibitemShut {NoStop}%
\bibitem [{\citenamefont {Monkman}\ and\ \citenamefont {Sirker}(2023)}]{monkman2023}%
  \BibitemOpen
  \bibfield  {author} {\bibinfo {author} {\bibfnamefont {K.}~\bibnamefont {Monkman}}\ and\ \bibinfo {author} {\bibfnamefont {J.}~\bibnamefont {Sirker}},\ }\bibfield  {title} {\bibinfo {title} {{Symmetry-resolved entanglement: general considerations, calculation from correlation functions, and bounds for symmetry-protected topological phases}},\ }\href {https://doi.org/10.1088/1751-8121/ad086d} {\bibfield  {journal} {\bibinfo  {journal} {Journal of Physics A: Mathematical and Theoretical}\ }\textbf {\bibinfo {volume} {56}},\ \bibinfo {pages} {495001} (\bibinfo {year} {2023})}\BibitemShut {NoStop}%
\bibitem [{\citenamefont {Berthiere}\ and\ \citenamefont {Parez}(2023)}]{berthiere2023}%
  \BibitemOpen
  \bibfield  {author} {\bibinfo {author} {\bibfnamefont {C.}~\bibnamefont {Berthiere}}\ and\ \bibinfo {author} {\bibfnamefont {G.}~\bibnamefont {Parez}},\ }\href {https://arxiv.org/abs/2307.11009} {\bibinfo {title} {{Reflected entropy and computable cross-norm negativity: Free theories and symmetry resolution}}} (\bibinfo {year} {2023}),\ \Eprint {https://arxiv.org/abs/2307.11009} {arXiv:2307.11009 [hep-th]} \BibitemShut {NoStop}%
\bibitem [{\citenamefont {Meyer}\ and\ \citenamefont {Wallach}(2002)}]{meyer2002}%
  \BibitemOpen
  \bibfield  {author} {\bibinfo {author} {\bibfnamefont {D.~A.}\ \bibnamefont {Meyer}}\ and\ \bibinfo {author} {\bibfnamefont {N.~R.}\ \bibnamefont {Wallach}},\ }\bibfield  {title} {\bibinfo {title} {{Global entanglement in multiparticle systems}},\ }\href {https://doi.org/10.1063/1.1497700} {\bibfield  {journal} {\bibinfo  {journal} {Journal of Mathematical Physics}\ }\textbf {\bibinfo {volume} {43}},\ \bibinfo {pages} {4273} (\bibinfo {year} {2002})}\BibitemShut {NoStop}%
\bibitem [{\citenamefont {Wootters}(2001)}]{wootters2001}%
  \BibitemOpen
  \bibfield  {author} {\bibinfo {author} {\bibfnamefont {W.~K.}\ \bibnamefont {Wootters}},\ }\bibfield  {title} {\bibinfo {title} {{Entanglement of formation and concurrence.}},\ }\href {https://doi.org/10.26421/qic1.1-3} {\bibfield  {journal} {\bibinfo  {journal} {Quantum Inf. Comput.}\ }\textbf {\bibinfo {volume} {1}},\ \bibinfo {pages} {27} (\bibinfo {year} {2001})}\BibitemShut {NoStop}%
\bibitem [{\citenamefont {Carisch}\ and\ \citenamefont {Zilberberg}(2023)}]{carisch2023}%
  \BibitemOpen
  \bibfield  {author} {\bibinfo {author} {\bibfnamefont {C.}~\bibnamefont {Carisch}}\ and\ \bibinfo {author} {\bibfnamefont {O.}~\bibnamefont {Zilberberg}},\ }\bibfield  {title} {\bibinfo {title} {{Efficient separation of quantum from classical correlations for mixed states with a fixed charge}},\ }\href {https://doi.org/10.22331/q-2023-03-20-954} {\bibfield  {journal} {\bibinfo  {journal} {Quantum}\ }\textbf {\bibinfo {volume} {7}},\ \bibinfo {pages} {954} (\bibinfo {year} {2023})}\BibitemShut {NoStop}%
\bibitem [{\citenamefont {Gurvits}(2003)}]{gurvits2003}%
  \BibitemOpen
  \bibfield  {author} {\bibinfo {author} {\bibfnamefont {L.}~\bibnamefont {Gurvits}},\ }\href {https://arxiv.org/abs/quant-ph/0303055} {\bibinfo {title} {{Classical deterministic complexity of Edmonds' problem and Quantum Entanglement}}} (\bibinfo {year} {2003}),\ \Eprint {https://arxiv.org/abs/quant-ph/0303055} {arXiv:quant-ph/0303055 [quant-ph]} \BibitemShut {NoStop}%
\bibitem [{\citenamefont {Gharibian}(2009)}]{gharibian2009}%
  \BibitemOpen
  \bibfield  {author} {\bibinfo {author} {\bibfnamefont {S.}~\bibnamefont {Gharibian}},\ }\href {https://arxiv.org/abs/0810.4507} {\bibinfo {title} {{Strong NP-Hardness of the Quantum Separability Problem}}} (\bibinfo {year} {2009}),\ \Eprint {https://arxiv.org/abs/0810.4507} {arXiv:0810.4507 [quant-ph]} \BibitemShut {NoStop}%
\bibitem [{\citenamefont {Goldstein}\ and\ \citenamefont {Sela}(2018)}]{goldstein2018}%
  \BibitemOpen
  \bibfield  {author} {\bibinfo {author} {\bibfnamefont {M.}~\bibnamefont {Goldstein}}\ and\ \bibinfo {author} {\bibfnamefont {E.}~\bibnamefont {Sela}},\ }\bibfield  {title} {\bibinfo {title} {{Symmetry-Resolved Entanglement in Many-Body Systems}},\ }\href {https://doi.org/10.1103/PhysRevLett.120.200602} {\bibfield  {journal} {\bibinfo  {journal} {Phys. Rev. Lett.}\ }\textbf {\bibinfo {volume} {120}},\ \bibinfo {pages} {200602} (\bibinfo {year} {2018})}\BibitemShut {NoStop}%
\bibitem [{\citenamefont {Estienne}\ \emph {et~al.}(2021)\citenamefont {Estienne}, \citenamefont {Ikhlef},\ and\ \citenamefont {Morin-Duchesne}}]{estienne2021}%
  \BibitemOpen
  \bibfield  {author} {\bibinfo {author} {\bibfnamefont {B.}~\bibnamefont {Estienne}}, \bibinfo {author} {\bibfnamefont {Y.}~\bibnamefont {Ikhlef}},\ and\ \bibinfo {author} {\bibfnamefont {A.}~\bibnamefont {Morin-Duchesne}},\ }\bibfield  {title} {\bibinfo {title} {{Finite-size corrections in critical symmetry-resolved entanglement}},\ }\bibfield  {journal} {\bibinfo  {journal} {SciPost Physics}\ }\textbf {\bibinfo {volume} {10}},\ \href {https://doi.org/10.21468/scipostphys.10.3.054} {10.21468/scipostphys.10.3.054} (\bibinfo {year} {2021})\BibitemShut {NoStop}%
\bibitem [{\citenamefont {Macieszczak}\ \emph {et~al.}(2019)\citenamefont {Macieszczak}, \citenamefont {Levi}, \citenamefont {Macr\`{\i}}, \citenamefont {Lesanovsky},\ and\ \citenamefont {Garrahan}}]{macieszczak2019}%
  \BibitemOpen
  \bibfield  {author} {\bibinfo {author} {\bibfnamefont {K.}~\bibnamefont {Macieszczak}}, \bibinfo {author} {\bibfnamefont {E.}~\bibnamefont {Levi}}, \bibinfo {author} {\bibfnamefont {T.}~\bibnamefont {Macr\`{\i}}}, \bibinfo {author} {\bibfnamefont {I.}~\bibnamefont {Lesanovsky}},\ and\ \bibinfo {author} {\bibfnamefont {J.~P.}\ \bibnamefont {Garrahan}},\ }\bibfield  {title} {\bibinfo {title} {{Coherence, entanglement, and quantumness in closed and open systems with conserved charge, with an application to many-body localization}},\ }\href {https://doi.org/10.1103/PhysRevA.99.052354} {\bibfield  {journal} {\bibinfo  {journal} {Phys. Rev. A}\ }\textbf {\bibinfo {volume} {99}},\ \bibinfo {pages} {052354} (\bibinfo {year} {2019})}\BibitemShut {NoStop}%
\bibitem [{\citenamefont {Ma}\ \emph {et~al.}(2022)\citenamefont {Ma}, \citenamefont {Han}, \citenamefont {Meir},\ and\ \citenamefont {Sela}}]{ma2022}%
  \BibitemOpen
  \bibfield  {author} {\bibinfo {author} {\bibfnamefont {Z.}~\bibnamefont {Ma}}, \bibinfo {author} {\bibfnamefont {C.}~\bibnamefont {Han}}, \bibinfo {author} {\bibfnamefont {Y.}~\bibnamefont {Meir}},\ and\ \bibinfo {author} {\bibfnamefont {E.}~\bibnamefont {Sela}},\ }\bibfield  {title} {\bibinfo {title} {{Symmetric inseparability and number entanglement in charge-conserving mixed states}},\ }\bibfield  {journal} {\bibinfo  {journal} {Physical Review A}\ }\textbf {\bibinfo {volume} {105}},\ \href {https://doi.org/10.1103/physreva.105.042416} {10.1103/physreva.105.042416} (\bibinfo {year} {2022})\BibitemShut {NoStop}%
\bibitem [{\citenamefont {Bonsignori}\ \emph {et~al.}(2019)\citenamefont {Bonsignori}, \citenamefont {Ruggiero},\ and\ \citenamefont {Calabrese}}]{bonsignori2019}%
  \BibitemOpen
  \bibfield  {author} {\bibinfo {author} {\bibfnamefont {R.}~\bibnamefont {Bonsignori}}, \bibinfo {author} {\bibfnamefont {P.}~\bibnamefont {Ruggiero}},\ and\ \bibinfo {author} {\bibfnamefont {P.}~\bibnamefont {Calabrese}},\ }\bibfield  {title} {\bibinfo {title} {{Symmetry resolved entanglement in free fermionic systems}},\ }\href {https://doi.org/10.1088/1751-8121/ab4b77} {\bibfield  {journal} {\bibinfo  {journal} {Journal of Physics A: Mathematical and Theoretical}\ }\textbf {\bibinfo {volume} {52}},\ \bibinfo {pages} {475302} (\bibinfo {year} {2019})}\BibitemShut {NoStop}%
\bibitem [{\citenamefont {Xavier}\ \emph {et~al.}(2018)\citenamefont {Xavier}, \citenamefont {Alcaraz},\ and\ \citenamefont {Sierra}}]{xavier2018}%
  \BibitemOpen
  \bibfield  {author} {\bibinfo {author} {\bibfnamefont {J.~C.}\ \bibnamefont {Xavier}}, \bibinfo {author} {\bibfnamefont {F.~C.}\ \bibnamefont {Alcaraz}},\ and\ \bibinfo {author} {\bibfnamefont {G.}~\bibnamefont {Sierra}},\ }\bibfield  {title} {\bibinfo {title} {{Equipartition of the entanglement entropy}},\ }\href {https://doi.org/10.1103/PhysRevB.98.041106} {\bibfield  {journal} {\bibinfo  {journal} {Phys. Rev. B}\ }\textbf {\bibinfo {volume} {98}},\ \bibinfo {pages} {041106} (\bibinfo {year} {2018})}\BibitemShut {NoStop}%
\bibitem [{\citenamefont {Piroli}\ \emph {et~al.}(2022)\citenamefont {Piroli}, \citenamefont {Vernier}, \citenamefont {Collura},\ and\ \citenamefont {Calabrese}}]{piroli2022}%
  \BibitemOpen
  \bibfield  {author} {\bibinfo {author} {\bibfnamefont {L.}~\bibnamefont {Piroli}}, \bibinfo {author} {\bibfnamefont {E.}~\bibnamefont {Vernier}}, \bibinfo {author} {\bibfnamefont {M.}~\bibnamefont {Collura}},\ and\ \bibinfo {author} {\bibfnamefont {P.}~\bibnamefont {Calabrese}},\ }\bibfield  {title} {\bibinfo {title} {{Thermodynamic symmetry resolved entanglement entropies in integrable systems}},\ }\href {https://doi.org/10.1088/1742-5468/ac7a2d} {\bibfield  {journal} {\bibinfo  {journal} {Journal of Statistical Mechanics: Theory and Experiment}\ }\textbf {\bibinfo {volume} {2022}},\ \bibinfo {pages} {073102} (\bibinfo {year} {2022})}\BibitemShut {NoStop}%
\bibitem [{\citenamefont {Jones}(2022)}]{jones2022}%
  \BibitemOpen
  \bibfield  {author} {\bibinfo {author} {\bibfnamefont {N.~G.}\ \bibnamefont {Jones}},\ }\bibfield  {title} {\bibinfo {title} {{Symmetry-Resolved Entanglement Entropy in Critical Free-Fermion Chains}},\ }\bibfield  {journal} {\bibinfo  {journal} {Journal of Statistical Physics}\ }\textbf {\bibinfo {volume} {188}},\ \href {https://doi.org/10.1007/s10955-022-02941-3} {10.1007/s10955-022-02941-3} (\bibinfo {year} {2022})\BibitemShut {NoStop}%
\bibitem [{\citenamefont {Bianchi}\ \emph {et~al.}(2024)\citenamefont {Bianchi}, \citenamefont {Dona},\ and\ \citenamefont {Kumar}}]{bianchi2024}%
  \BibitemOpen
  \bibfield  {author} {\bibinfo {author} {\bibfnamefont {E.}~\bibnamefont {Bianchi}}, \bibinfo {author} {\bibfnamefont {P.}~\bibnamefont {Dona}},\ and\ \bibinfo {author} {\bibfnamefont {R.}~\bibnamefont {Kumar}},\ }\bibfield  {title} {\bibinfo {title} {{Non-Abelian symmetry-resolved entanglement entropy}},\ }\bibfield  {journal} {\bibinfo  {journal} {SciPost Physics}\ }\textbf {\bibinfo {volume} {17}},\ \href {https://doi.org/10.21468/scipostphys.17.5.127} {10.21468/scipostphys.17.5.127} (\bibinfo {year} {2024})\BibitemShut {NoStop}%
\bibitem [{\citenamefont {Calabrese}\ \emph {et~al.}(2021)\citenamefont {Calabrese}, \citenamefont {Dubail},\ and\ \citenamefont {Murciano}}]{calabrese2021}%
  \BibitemOpen
  \bibfield  {author} {\bibinfo {author} {\bibfnamefont {P.}~\bibnamefont {Calabrese}}, \bibinfo {author} {\bibfnamefont {J.}~\bibnamefont {Dubail}},\ and\ \bibinfo {author} {\bibfnamefont {S.}~\bibnamefont {Murciano}},\ }\bibfield  {title} {\bibinfo {title} {{Symmetry-resolved entanglement entropy in Wess-Zumino-Witten models}},\ }\href {https://doi.org/10.1007/jhep10(2021)067} {\bibfield  {journal} {\bibinfo  {journal} {Journal of High Energy Physics}\ }\textbf {\bibinfo {volume} {2021}},\ \bibinfo {pages} {1} (\bibinfo {year} {2021})}\BibitemShut {NoStop}%
\bibitem [{\citenamefont {Neven}\ \emph {et~al.}(2021)\citenamefont {Neven}, \citenamefont {Carrasco}, \citenamefont {Vitale}, \citenamefont {Kokail}, \citenamefont {Elben}, \citenamefont {Dalmonte}, \citenamefont {Calabrese}, \citenamefont {Zoller}, \citenamefont {Vermersch}, \citenamefont {Kueng} \emph {et~al.}}]{neven2021}%
  \BibitemOpen
  \bibfield  {author} {\bibinfo {author} {\bibfnamefont {A.}~\bibnamefont {Neven}}, \bibinfo {author} {\bibfnamefont {J.}~\bibnamefont {Carrasco}}, \bibinfo {author} {\bibfnamefont {V.}~\bibnamefont {Vitale}}, \bibinfo {author} {\bibfnamefont {C.}~\bibnamefont {Kokail}}, \bibinfo {author} {\bibfnamefont {A.}~\bibnamefont {Elben}}, \bibinfo {author} {\bibfnamefont {M.}~\bibnamefont {Dalmonte}}, \bibinfo {author} {\bibfnamefont {P.}~\bibnamefont {Calabrese}}, \bibinfo {author} {\bibfnamefont {P.}~\bibnamefont {Zoller}}, \bibinfo {author} {\bibfnamefont {B.}~\bibnamefont {Vermersch}}, \bibinfo {author} {\bibfnamefont {R.}~\bibnamefont {Kueng}}, \emph {et~al.},\ }\bibfield  {title} {\bibinfo {title} {{Symmetry-resolved entanglement detection using partial transpose moments}},\ }\href {https://doi.org/10.1038/s41534-021-00487-y} {\bibfield  {journal} {\bibinfo  {journal} {npj Quantum Information}\ }\textbf {\bibinfo {volume} {7}},\ \bibinfo {pages} {152} (\bibinfo {year} {2021})}\BibitemShut {NoStop}%
\bibitem [{\citenamefont {Kusuki}\ \emph {et~al.}(2023)\citenamefont {Kusuki}, \citenamefont {Murciano}, \citenamefont {Ooguri},\ and\ \citenamefont {Pal}}]{kusuki2023}%
  \BibitemOpen
  \bibfield  {author} {\bibinfo {author} {\bibfnamefont {Y.}~\bibnamefont {Kusuki}}, \bibinfo {author} {\bibfnamefont {S.}~\bibnamefont {Murciano}}, \bibinfo {author} {\bibfnamefont {H.}~\bibnamefont {Ooguri}},\ and\ \bibinfo {author} {\bibfnamefont {S.}~\bibnamefont {Pal}},\ }\bibfield  {title} {\bibinfo {title} {{Symmetry-resolved entanglement entropy, spectra \& boundary conformal field theory}},\ }\href {https://doi.org/10.1007/jhep11(2023)216} {\bibfield  {journal} {\bibinfo  {journal} {Journal of High Energy Physics}\ }\textbf {\bibinfo {volume} {2023}},\ \bibinfo {pages} {1} (\bibinfo {year} {2023})}\BibitemShut {NoStop}%
\bibitem [{\citenamefont {Hastings}(2007)}]{hastings2007}%
  \BibitemOpen
  \bibfield  {author} {\bibinfo {author} {\bibfnamefont {M.~B.}\ \bibnamefont {Hastings}},\ }\bibfield  {title} {\bibinfo {title} {{An area law for one-dimensional quantum systems}},\ }\href {https://doi.org/10.1088/1742-5468/2007/08/p08024} {\bibfield  {journal} {\bibinfo  {journal} {Journal of statistical mechanics: theory and experiment}\ }\textbf {\bibinfo {volume} {2007}},\ \bibinfo {pages} {P08024} (\bibinfo {year} {2007})}\BibitemShut {NoStop}%
\bibitem [{\citenamefont {Vidal}\ \emph {et~al.}(2003)\citenamefont {Vidal}, \citenamefont {Latorre}, \citenamefont {Rico},\ and\ \citenamefont {Kitaev}}]{vidal2003}%
  \BibitemOpen
  \bibfield  {author} {\bibinfo {author} {\bibfnamefont {G.}~\bibnamefont {Vidal}}, \bibinfo {author} {\bibfnamefont {J.~I.}\ \bibnamefont {Latorre}}, \bibinfo {author} {\bibfnamefont {E.}~\bibnamefont {Rico}},\ and\ \bibinfo {author} {\bibfnamefont {A.}~\bibnamefont {Kitaev}},\ }\bibfield  {title} {\bibinfo {title} {{Entanglement in Quantum Critical Phenomena}},\ }\href {https://doi.org/10.1103/PhysRevLett.90.227902} {\bibfield  {journal} {\bibinfo  {journal} {Phys. Rev. Lett.}\ }\textbf {\bibinfo {volume} {90}},\ \bibinfo {pages} {227902} (\bibinfo {year} {2003})}\BibitemShut {NoStop}%
\bibitem [{\citenamefont {Eisert}\ \emph {et~al.}(2010)\citenamefont {Eisert}, \citenamefont {Cramer},\ and\ \citenamefont {Plenio}}]{eisert2010}%
  \BibitemOpen
  \bibfield  {author} {\bibinfo {author} {\bibfnamefont {J.}~\bibnamefont {Eisert}}, \bibinfo {author} {\bibfnamefont {M.}~\bibnamefont {Cramer}},\ and\ \bibinfo {author} {\bibfnamefont {M.~B.}\ \bibnamefont {Plenio}},\ }\bibfield  {title} {\bibinfo {title} {{Colloquium: Area laws for the entanglement entropy}},\ }\href {https://doi.org/10.1103/revmodphys.82.277} {\bibfield  {journal} {\bibinfo  {journal} {Reviews of Modern Physics}\ }\textbf {\bibinfo {volume} {82}},\ \bibinfo {pages} {277–306} (\bibinfo {year} {2010})}\BibitemShut {NoStop}%
\bibitem [{\citenamefont {Ares}\ \emph {et~al.}(2022)\citenamefont {Ares}, \citenamefont {Murciano},\ and\ \citenamefont {Calabrese}}]{ares2022}%
  \BibitemOpen
  \bibfield  {author} {\bibinfo {author} {\bibfnamefont {F.}~\bibnamefont {Ares}}, \bibinfo {author} {\bibfnamefont {S.}~\bibnamefont {Murciano}},\ and\ \bibinfo {author} {\bibfnamefont {P.}~\bibnamefont {Calabrese}},\ }\bibfield  {title} {\bibinfo {title} {{Symmetry-resolved entanglement in a long-range free-fermion chain}},\ }\href {https://doi.org/10.1088/1742-5468/ac7644} {\bibfield  {journal} {\bibinfo  {journal} {Journal of Statistical Mechanics: Theory and Experiment}\ }\textbf {\bibinfo {volume} {2022}},\ \bibinfo {pages} {063104} (\bibinfo {year} {2022})}\BibitemShut {NoStop}%
\bibitem [{\citenamefont {Parez}\ \emph {et~al.}(2021)\citenamefont {Parez}, \citenamefont {Bonsignori},\ and\ \citenamefont {Calabrese}}]{parez2021}%
  \BibitemOpen
  \bibfield  {author} {\bibinfo {author} {\bibfnamefont {G.}~\bibnamefont {Parez}}, \bibinfo {author} {\bibfnamefont {R.}~\bibnamefont {Bonsignori}},\ and\ \bibinfo {author} {\bibfnamefont {P.}~\bibnamefont {Calabrese}},\ }\bibfield  {title} {\bibinfo {title} {{Exact quench dynamics of symmetry resolved entanglement in a free fermion chain}},\ }\href {https://doi.org/10.1088/1742-5468/ac21d7} {\bibfield  {journal} {\bibinfo  {journal} {Journal of Statistical Mechanics: Theory and Experiment}\ }\textbf {\bibinfo {volume} {2021}},\ \bibinfo {pages} {093102} (\bibinfo {year} {2021})}\BibitemShut {NoStop}%
\bibitem [{\citenamefont {Osterloh}\ \emph {et~al.}(2002)\citenamefont {Osterloh}, \citenamefont {Amico}, \citenamefont {Falci},\ and\ \citenamefont {Fazio}}]{osterloh2002}%
  \BibitemOpen
  \bibfield  {author} {\bibinfo {author} {\bibfnamefont {A.}~\bibnamefont {Osterloh}}, \bibinfo {author} {\bibfnamefont {L.}~\bibnamefont {Amico}}, \bibinfo {author} {\bibfnamefont {G.}~\bibnamefont {Falci}},\ and\ \bibinfo {author} {\bibfnamefont {R.}~\bibnamefont {Fazio}},\ }\bibfield  {title} {\bibinfo {title} {{Scaling of entanglement close to a quantum phase transition}},\ }\href {https://doi.org/10.1038/416608a} {\bibfield  {journal} {\bibinfo  {journal} {Nature}\ }\textbf {\bibinfo {volume} {416}},\ \bibinfo {pages} {608–610} (\bibinfo {year} {2002})}\BibitemShut {NoStop}%
\bibitem [{\citenamefont {Osborne}\ and\ \citenamefont {Nielsen}(2002)}]{osborne2002}%
  \BibitemOpen
  \bibfield  {author} {\bibinfo {author} {\bibfnamefont {T.~J.}\ \bibnamefont {Osborne}}\ and\ \bibinfo {author} {\bibfnamefont {M.~A.}\ \bibnamefont {Nielsen}},\ }\bibfield  {title} {\bibinfo {title} {{Entanglement in a simple quantum phase transition}},\ }\href {https://doi.org/10.1103/PhysRevA.66.032110} {\bibfield  {journal} {\bibinfo  {journal} {Phys. Rev. A}\ }\textbf {\bibinfo {volume} {66}},\ \bibinfo {pages} {032110} (\bibinfo {year} {2002})}\BibitemShut {NoStop}%
\bibitem [{\citenamefont {Li}\ \emph {et~al.}(2024)\citenamefont {Li}, \citenamefont {Zhou}, \citenamefont {Zhang}, \citenamefont {Bai},\ and\ \citenamefont {Lin}}]{li2024}%
  \BibitemOpen
  \bibfield  {author} {\bibinfo {author} {\bibfnamefont {Y.~C.}\ \bibnamefont {Li}}, \bibinfo {author} {\bibfnamefont {Y.~H.}\ \bibnamefont {Zhou}}, \bibinfo {author} {\bibfnamefont {Y.}~\bibnamefont {Zhang}}, \bibinfo {author} {\bibfnamefont {Y.~K.}\ \bibnamefont {Bai}},\ and\ \bibinfo {author} {\bibfnamefont {H.~Q.}\ \bibnamefont {Lin}},\ }\href {https://arxiv.org/abs/2401.15593} {\bibinfo {title} {{Multipartite entanglement serves as a faithful detector for quantum phase transitions}}} (\bibinfo {year} {2024}),\ \Eprint {https://arxiv.org/abs/2401.15593} {arXiv:2401.15593 [quant-ph]} \BibitemShut {NoStop}%
\bibitem [{\citenamefont {Iftikhar}\ \emph {et~al.}(2021)\citenamefont {Iftikhar}, \citenamefont {Usman},\ and\ \citenamefont {Khan}}]{iftikhar2021}%
  \BibitemOpen
  \bibfield  {author} {\bibinfo {author} {\bibfnamefont {M.~T.}\ \bibnamefont {Iftikhar}}, \bibinfo {author} {\bibfnamefont {M.}~\bibnamefont {Usman}},\ and\ \bibinfo {author} {\bibfnamefont {K.}~\bibnamefont {Khan}},\ }\bibfield  {title} {\bibinfo {title} {{Multipartite entanglement and criticality in two-dimensional XXZ model}},\ }\href {https://doi.org/10.1007/s11128-021-03185-y} {\bibfield  {journal} {\bibinfo  {journal} {Quantum Information Processing}\ }\textbf {\bibinfo {volume} {20}},\ \bibinfo {pages} {259} (\bibinfo {year} {2021})}\BibitemShut {NoStop}%
\bibitem [{\citenamefont {Giampaolo}\ and\ \citenamefont {Hiesmayr}(2013)}]{giampaolo2013}%
  \BibitemOpen
  \bibfield  {author} {\bibinfo {author} {\bibfnamefont {S.}~\bibnamefont {Giampaolo}}\ and\ \bibinfo {author} {\bibfnamefont {B.}~\bibnamefont {Hiesmayr}},\ }\bibfield  {title} {\bibinfo {title} {{Genuine multipartite entanglement in the XY model}},\ }\href {https://doi.org/10.1088/1742-6596/2667/1/012034} {\bibfield  {journal} {\bibinfo  {journal} {Physical Review A—Atomic, Molecular, and Optical Physics}\ }\textbf {\bibinfo {volume} {88}},\ \bibinfo {pages} {052305} (\bibinfo {year} {2013})}\BibitemShut {NoStop}%
\bibitem [{\citenamefont {Montakhab}\ and\ \citenamefont {Asadian}(2010)}]{montakhab2010}%
  \BibitemOpen
  \bibfield  {author} {\bibinfo {author} {\bibfnamefont {A.}~\bibnamefont {Montakhab}}\ and\ \bibinfo {author} {\bibfnamefont {A.}~\bibnamefont {Asadian}},\ }\bibfield  {title} {\bibinfo {title} {{Multipartite entanglement and quantum phase transitions in the one-, two-, and three-dimensional transverse-field Ising model}},\ }\href {https://doi.org/10.1103/physreva.82.062313} {\bibfield  {journal} {\bibinfo  {journal} {Physical Review A—Atomic, Molecular, and Optical Physics}\ }\textbf {\bibinfo {volume} {82}},\ \bibinfo {pages} {062313} (\bibinfo {year} {2010})}\BibitemShut {NoStop}%
\bibitem [{\citenamefont {Hofmann}\ \emph {et~al.}(2014)\citenamefont {Hofmann}, \citenamefont {Osterloh},\ and\ \citenamefont {G{\"u}hne}}]{hofmann2014}%
  \BibitemOpen
  \bibfield  {author} {\bibinfo {author} {\bibfnamefont {M.}~\bibnamefont {Hofmann}}, \bibinfo {author} {\bibfnamefont {A.}~\bibnamefont {Osterloh}},\ and\ \bibinfo {author} {\bibfnamefont {O.}~\bibnamefont {G{\"u}hne}},\ }\bibfield  {title} {\bibinfo {title} {{Scaling of genuine multiparticle entanglement close to a quantum phase transition}},\ }\href {https://doi.org/10.1103/physrevb.89.134101} {\bibfield  {journal} {\bibinfo  {journal} {Physical Review B}\ }\textbf {\bibinfo {volume} {89}},\ \bibinfo {pages} {134101} (\bibinfo {year} {2014})}\BibitemShut {NoStop}%
\bibitem [{\citenamefont {Sun}\ \emph {et~al.}(2024)\citenamefont {Sun}, \citenamefont {Ge},\ and\ \citenamefont {Fan}}]{sun2024}%
  \BibitemOpen
  \bibfield  {author} {\bibinfo {author} {\bibfnamefont {H.-Y.}\ \bibnamefont {Sun}}, \bibinfo {author} {\bibfnamefont {Z.-Y.}\ \bibnamefont {Ge}},\ and\ \bibinfo {author} {\bibfnamefont {H.}~\bibnamefont {Fan}},\ }\bibfield  {title} {\bibinfo {title} {{Multipartite entanglement in crossing the quantum critical point}},\ }\href {https://doi.org/10.1088/1402-4896/ad8e0d} {\bibfield  {journal} {\bibinfo  {journal} {Physica Scripta}\ }\textbf {\bibinfo {volume} {99}},\ \bibinfo {pages} {125111} (\bibinfo {year} {2024})}\BibitemShut {NoStop}%
\bibitem [{\citenamefont {Brennen}(2003)}]{brennen2003}%
  \BibitemOpen
  \bibfield  {author} {\bibinfo {author} {\bibfnamefont {G.~K.}\ \bibnamefont {Brennen}},\ }\href {https://arxiv.org/abs/quant-ph/0305094} {\bibinfo {title} {{An observable measure of entanglement for pure states of multi-qubit systems}}} (\bibinfo {year} {2003}),\ \Eprint {https://arxiv.org/abs/quant-ph/0305094} {arXiv:quant-ph/0305094 [quant-ph]} \BibitemShut {NoStop}%
\bibitem [{\citenamefont {Xin}\ \emph {et~al.}(2005)\citenamefont {Xin}, \citenamefont {Song},\ and\ \citenamefont {Sun}}]{xin2005}%
  \BibitemOpen
  \bibfield  {author} {\bibinfo {author} {\bibfnamefont {R.}~\bibnamefont {Xin}}, \bibinfo {author} {\bibfnamefont {Z.}~\bibnamefont {Song}},\ and\ \bibinfo {author} {\bibfnamefont {C.}~\bibnamefont {Sun}},\ }\bibfield  {title} {\bibinfo {title} {{Engineering antiferromagnetic Heisenberg spin chains for maximizing of the groundstate entanglement}},\ }\href {https://doi.org/10.1016/j.physleta.2005.05.007} {\bibfield  {journal} {\bibinfo  {journal} {Physics Letters A}\ }\textbf {\bibinfo {volume} {342}},\ \bibinfo {pages} {30} (\bibinfo {year} {2005})}\BibitemShut {NoStop}%
\bibitem [{\citenamefont {Wang}\ \emph {et~al.}(2010)\citenamefont {Wang}, \citenamefont {Bayat}, \citenamefont {Schirmer},\ and\ \citenamefont {Bose}}]{wang2010}%
  \BibitemOpen
  \bibfield  {author} {\bibinfo {author} {\bibfnamefont {X.}~\bibnamefont {Wang}}, \bibinfo {author} {\bibfnamefont {A.}~\bibnamefont {Bayat}}, \bibinfo {author} {\bibfnamefont {S.~G.}\ \bibnamefont {Schirmer}},\ and\ \bibinfo {author} {\bibfnamefont {S.}~\bibnamefont {Bose}},\ }\bibfield  {title} {\bibinfo {title} {{Robust entanglement in antiferromagnetic Heisenberg chains by single-spin optimal control}},\ }\href {https://doi.org/10.1103/PhysRevA.81.032312} {\bibfield  {journal} {\bibinfo  {journal} {Phys. Rev. A}\ }\textbf {\bibinfo {volume} {81}},\ \bibinfo {pages} {032312} (\bibinfo {year} {2010})}\BibitemShut {NoStop}%
\bibitem [{\citenamefont {de~Oliveira}\ \emph {et~al.}(2006)\citenamefont {de~Oliveira}, \citenamefont {Rigolin},\ and\ \citenamefont {de~Oliveira}}]{deOliveira2006a}%
  \BibitemOpen
  \bibfield  {author} {\bibinfo {author} {\bibfnamefont {T.~R.}\ \bibnamefont {de~Oliveira}}, \bibinfo {author} {\bibfnamefont {G.}~\bibnamefont {Rigolin}},\ and\ \bibinfo {author} {\bibfnamefont {M.~C.}\ \bibnamefont {de~Oliveira}},\ }\bibfield  {title} {\bibinfo {title} {{Genuine multipartite entanglement in quantum phase transitions}},\ }\href {https://doi.org/10.1103/PhysRevA.73.010305} {\bibfield  {journal} {\bibinfo  {journal} {Phys. Rev. A}\ }\textbf {\bibinfo {volume} {73}},\ \bibinfo {pages} {010305} (\bibinfo {year} {2006})}\BibitemShut {NoStop}%
\bibitem [{\citenamefont {Jin}\ and\ \citenamefont {Korepin}(2004)}]{jin2004}%
  \BibitemOpen
  \bibfield  {author} {\bibinfo {author} {\bibfnamefont {B.-Q.}\ \bibnamefont {Jin}}\ and\ \bibinfo {author} {\bibfnamefont {V.~E.}\ \bibnamefont {Korepin}},\ }\bibfield  {title} {\bibinfo {title} {{Quantum Spin Chain, Toeplitz Determinants and the Fisher–Hartwig Conjecture}},\ }\href {https://doi.org/10.1023/b:joss.0000037230.37166.42} {\bibfield  {journal} {\bibinfo  {journal} {Journal of Statistical Physics}\ }\textbf {\bibinfo {volume} {116}},\ \bibinfo {pages} {79–95} (\bibinfo {year} {2004})}\BibitemShut {NoStop}%
\bibitem [{\citenamefont {Keating}\ and\ \citenamefont {Mezzadri}(2005)}]{keating2005}%
  \BibitemOpen
  \bibfield  {author} {\bibinfo {author} {\bibfnamefont {J.~P.}\ \bibnamefont {Keating}}\ and\ \bibinfo {author} {\bibfnamefont {F.}~\bibnamefont {Mezzadri}},\ }\bibfield  {title} {\bibinfo {title} {{Entanglement in Quantum Spin Chains, Symmetry Classes of Random Matrices, and Conformal Field Theory}},\ }\href {https://doi.org/10.1103/PhysRevLett.94.050501} {\bibfield  {journal} {\bibinfo  {journal} {Phys. Rev. Lett.}\ }\textbf {\bibinfo {volume} {94}},\ \bibinfo {pages} {050501} (\bibinfo {year} {2005})}\BibitemShut {NoStop}%
\bibitem [{\citenamefont {Latorre}\ \emph {et~al.}(2004)\citenamefont {Latorre}, \citenamefont {Rico},\ and\ \citenamefont {Vidal}}]{latorre2004}%
  \BibitemOpen
  \bibfield  {author} {\bibinfo {author} {\bibfnamefont {J.~I.}\ \bibnamefont {Latorre}}, \bibinfo {author} {\bibfnamefont {E.}~\bibnamefont {Rico}},\ and\ \bibinfo {author} {\bibfnamefont {G.}~\bibnamefont {Vidal}},\ }\href {https://arxiv.org/abs/quant-ph/0304098} {\bibinfo {title} {{Ground state entanglement in quantum spin chains}}} (\bibinfo {year} {2004}),\ \Eprint {https://arxiv.org/abs/quant-ph/0304098} {arXiv:quant-ph/0304098 [quant-ph]} \BibitemShut {NoStop}%
\bibitem [{\citenamefont {Barthel}\ \emph {et~al.}(2006)\citenamefont {Barthel}, \citenamefont {Dusuel},\ and\ \citenamefont {Vidal}}]{barthel2006}%
  \BibitemOpen
  \bibfield  {author} {\bibinfo {author} {\bibfnamefont {T.}~\bibnamefont {Barthel}}, \bibinfo {author} {\bibfnamefont {S.}~\bibnamefont {Dusuel}},\ and\ \bibinfo {author} {\bibfnamefont {J.}~\bibnamefont {Vidal}},\ }\bibfield  {title} {\bibinfo {title} {{Entanglement Entropy beyond the Free Case}},\ }\href {https://doi.org/10.1103/PhysRevLett.97.220402} {\bibfield  {journal} {\bibinfo  {journal} {Phys. Rev. Lett.}\ }\textbf {\bibinfo {volume} {97}},\ \bibinfo {pages} {220402} (\bibinfo {year} {2006})}\BibitemShut {NoStop}%
\bibitem [{\citenamefont {Or\'us}\ \emph {et~al.}(2006)\citenamefont {Or\'us}, \citenamefont {Latorre}, \citenamefont {Eisert},\ and\ \citenamefont {Cramer}}]{orus2018}%
  \BibitemOpen
  \bibfield  {author} {\bibinfo {author} {\bibfnamefont {R.}~\bibnamefont {Or\'us}}, \bibinfo {author} {\bibfnamefont {J.~I.}\ \bibnamefont {Latorre}}, \bibinfo {author} {\bibfnamefont {J.}~\bibnamefont {Eisert}},\ and\ \bibinfo {author} {\bibfnamefont {M.}~\bibnamefont {Cramer}},\ }\bibfield  {title} {\bibinfo {title} {{Half the entanglement in critical systems is distillable from a single specimen}},\ }\href {https://doi.org/10.1103/PhysRevA.73.060303} {\bibfield  {journal} {\bibinfo  {journal} {Phys. Rev. A}\ }\textbf {\bibinfo {volume} {73}},\ \bibinfo {pages} {060303} (\bibinfo {year} {2006})}\BibitemShut {NoStop}%
\bibitem [{\citenamefont {Tajik}\ \emph {et~al.}(2023)\citenamefont {Tajik}, \citenamefont {Kukuljan}, \citenamefont {Sotiriadis}, \citenamefont {Rauer}, \citenamefont {Schweigler}, \citenamefont {Cataldini}, \citenamefont {Sabino}, \citenamefont {M{\o}ller}, \citenamefont {Sch{\"u}ttelkopf}, \citenamefont {Ji} \emph {et~al.}}]{tajik2023}%
  \BibitemOpen
  \bibfield  {author} {\bibinfo {author} {\bibfnamefont {M.}~\bibnamefont {Tajik}}, \bibinfo {author} {\bibfnamefont {I.}~\bibnamefont {Kukuljan}}, \bibinfo {author} {\bibfnamefont {S.}~\bibnamefont {Sotiriadis}}, \bibinfo {author} {\bibfnamefont {B.}~\bibnamefont {Rauer}}, \bibinfo {author} {\bibfnamefont {T.}~\bibnamefont {Schweigler}}, \bibinfo {author} {\bibfnamefont {F.}~\bibnamefont {Cataldini}}, \bibinfo {author} {\bibfnamefont {J.}~\bibnamefont {Sabino}}, \bibinfo {author} {\bibfnamefont {F.}~\bibnamefont {M{\o}ller}}, \bibinfo {author} {\bibfnamefont {P.}~\bibnamefont {Sch{\"u}ttelkopf}}, \bibinfo {author} {\bibfnamefont {S.-C.}\ \bibnamefont {Ji}}, \emph {et~al.},\ }\bibfield  {title} {\bibinfo {title} {{Verification of the area law of mutual information in a quantum field simulator}},\ }\href {https://doi.org/10.1038/s41567-023-02027-1} {\bibfield  {journal} {\bibinfo  {journal} {Nature Physics}\ }\textbf {\bibinfo {volume} {19}},\ \bibinfo {pages} {1022} (\bibinfo {year} {2023})}\BibitemShut
  {NoStop}%
\bibitem [{\citenamefont {Herdman}\ \emph {et~al.}(2017)\citenamefont {Herdman}, \citenamefont {Roy}, \citenamefont {Melko},\ and\ \citenamefont {Maestro}}]{herdman2017}%
  \BibitemOpen
  \bibfield  {author} {\bibinfo {author} {\bibfnamefont {C.}~\bibnamefont {Herdman}}, \bibinfo {author} {\bibfnamefont {P.-N.}\ \bibnamefont {Roy}}, \bibinfo {author} {\bibfnamefont {R.}~\bibnamefont {Melko}},\ and\ \bibinfo {author} {\bibfnamefont {A.~D.}\ \bibnamefont {Maestro}},\ }\bibfield  {title} {\bibinfo {title} {{Entanglement area law in superfluid 4 He}},\ }\href {https://doi.org/10.1038/nphys4075} {\bibfield  {journal} {\bibinfo  {journal} {Nature Physics}\ }\textbf {\bibinfo {volume} {13}},\ \bibinfo {pages} {556} (\bibinfo {year} {2017})}\BibitemShut {NoStop}%
\bibitem [{\citenamefont {Lin}\ \emph {et~al.}(2024)\citenamefont {Lin}, \citenamefont {Zhou}, \citenamefont {Jiang}, \citenamefont {Wu}, \citenamefont {Chen}, \citenamefont {Liu}, \citenamefont {Wang}, \citenamefont {Ye},\ and\ \citenamefont {Jiang}}]{lin2024}%
  \BibitemOpen
  \bibfield  {author} {\bibinfo {author} {\bibfnamefont {Z.-K.}\ \bibnamefont {Lin}}, \bibinfo {author} {\bibfnamefont {Y.}~\bibnamefont {Zhou}}, \bibinfo {author} {\bibfnamefont {B.}~\bibnamefont {Jiang}}, \bibinfo {author} {\bibfnamefont {B.-Q.}\ \bibnamefont {Wu}}, \bibinfo {author} {\bibfnamefont {L.-M.}\ \bibnamefont {Chen}}, \bibinfo {author} {\bibfnamefont {X.-Y.}\ \bibnamefont {Liu}}, \bibinfo {author} {\bibfnamefont {L.-W.}\ \bibnamefont {Wang}}, \bibinfo {author} {\bibfnamefont {P.}~\bibnamefont {Ye}},\ and\ \bibinfo {author} {\bibfnamefont {J.-H.}\ \bibnamefont {Jiang}},\ }\bibfield  {title} {\bibinfo {title} {{Measuring entanglement entropy and its topological signature for phononic systems}},\ }\href {https://doi.org/10.1038/s41467-024-45887-8} {\bibfield  {journal} {\bibinfo  {journal} {Nature Communications}\ }\textbf {\bibinfo {volume} {15}},\ \bibinfo {pages} {1601} (\bibinfo {year} {2024})}\BibitemShut {NoStop}%
\bibitem [{\citenamefont {Low}\ and\ \citenamefont {Yin}(2024)}]{low2024}%
  \BibitemOpen
  \bibfield  {author} {\bibinfo {author} {\bibfnamefont {I.}~\bibnamefont {Low}}\ and\ \bibinfo {author} {\bibfnamefont {Z.}~\bibnamefont {Yin}},\ }\href {https://arxiv.org/abs/2405.08056} {\bibinfo {title} {{An Area Law for Entanglement Entropy in Particle Scattering}}} (\bibinfo {year} {2024}),\ \Eprint {https://arxiv.org/abs/2405.08056} {arXiv:2405.08056 [hep-th]} \BibitemShut {NoStop}%
\bibitem [{\citenamefont {De~Chiara}\ and\ \citenamefont {Sanpera}(2018)}]{deChiara2018}%
  \BibitemOpen
  \bibfield  {author} {\bibinfo {author} {\bibfnamefont {G.}~\bibnamefont {De~Chiara}}\ and\ \bibinfo {author} {\bibfnamefont {A.}~\bibnamefont {Sanpera}},\ }\bibfield  {title} {\bibinfo {title} {{Genuine quantum correlations in quantum many-body systems: a review of recent progress}},\ }\href {https://doi.org/10.1088/1361-6633/aabf61} {\bibfield  {journal} {\bibinfo  {journal} {Reports on Progress in Physics}\ }\textbf {\bibinfo {volume} {81}},\ \bibinfo {pages} {074002} (\bibinfo {year} {2018})}\BibitemShut {NoStop}%
\bibitem [{\citenamefont {Wu}\ \emph {et~al.}(2004)\citenamefont {Wu}, \citenamefont {Sarandy},\ and\ \citenamefont {Lidar}}]{wu2004}%
  \BibitemOpen
  \bibfield  {author} {\bibinfo {author} {\bibfnamefont {L.-A.}\ \bibnamefont {Wu}}, \bibinfo {author} {\bibfnamefont {M.~S.}\ \bibnamefont {Sarandy}},\ and\ \bibinfo {author} {\bibfnamefont {D.}~\bibnamefont {Lidar}},\ }\bibfield  {title} {\bibinfo {title} {{Quantum phase transitions and bipartite entanglement}},\ }\href {https://doi.org/10.1103/physrevlett.93.250404} {\bibfield  {journal} {\bibinfo  {journal} {Physical review letters}\ }\textbf {\bibinfo {volume} {93}},\ \bibinfo {pages} {250404} (\bibinfo {year} {2004})}\BibitemShut {NoStop}%
\bibitem [{\citenamefont {Yang}(2005)}]{yang2005}%
  \BibitemOpen
  \bibfield  {author} {\bibinfo {author} {\bibfnamefont {M.-F.}\ \bibnamefont {Yang}},\ }\bibfield  {title} {\bibinfo {title} {{Reexamination of entanglement and the quantum phase transition}},\ }\href {https://doi.org/10.1103/PhysRevA.71.030302} {\bibfield  {journal} {\bibinfo  {journal} {Phys. Rev. A}\ }\textbf {\bibinfo {volume} {71}},\ \bibinfo {pages} {030302} (\bibinfo {year} {2005})}\BibitemShut {NoStop}%
\bibitem [{\citenamefont {Zyczkowski}\ and\ \citenamefont {Bengtsson}(2006)}]{zyczkowski2006}%
  \BibitemOpen
  \bibfield  {author} {\bibinfo {author} {\bibfnamefont {K.}~\bibnamefont {Zyczkowski}}\ and\ \bibinfo {author} {\bibfnamefont {I.}~\bibnamefont {Bengtsson}},\ }\href {https://arxiv.org/abs/quant-ph/0606228} {\bibinfo {title} {{An Introduction to Quantum Entanglement: a Geometric Approach}}} (\bibinfo {year} {2006}),\ \Eprint {https://arxiv.org/abs/quant-ph/0606228} {arXiv:quant-ph/0606228 [quant-ph]} \BibitemShut {NoStop}%
\bibitem [{\citenamefont {Pauletti}\ \emph {et~al.}(2023)\citenamefont {Pauletti}, \citenamefont {Garcia}, \citenamefont {Canella},\ and\ \citenamefont {França}}]{pauletti2003}%
  \BibitemOpen
  \bibfield  {author} {\bibinfo {author} {\bibfnamefont {T.}~\bibnamefont {Pauletti}}, \bibinfo {author} {\bibfnamefont {M.}~\bibnamefont {Garcia}}, \bibinfo {author} {\bibfnamefont {G.~A.}\ \bibnamefont {Canella}},\ and\ \bibinfo {author} {\bibfnamefont {V.~V.}\ \bibnamefont {França}},\ }\href {https://arxiv.org/abs/2303.08075} {\bibinfo {title} {{Linear entropy fails to predict entanglement behavior in low-density fermionic systems}}} (\bibinfo {year} {2023}),\ \Eprint {https://arxiv.org/abs/2303.08075} {arXiv:2303.08075 [quant-ph]} \BibitemShut {NoStop}%
\bibitem [{\citenamefont {Scott}(2004)}]{scott2004}%
  \BibitemOpen
  \bibfield  {author} {\bibinfo {author} {\bibfnamefont {A.~J.}\ \bibnamefont {Scott}},\ }\bibfield  {title} {\bibinfo {title} {{Multipartite entanglement, quantum-error-correcting codes, and entangling power of quantum evolutions}},\ }\href {https://doi.org/10.1103/PhysRevA.69.052330} {\bibfield  {journal} {\bibinfo  {journal} {Phys. Rev. A}\ }\textbf {\bibinfo {volume} {69}},\ \bibinfo {pages} {052330} (\bibinfo {year} {2004})}\BibitemShut {NoStop}%
\bibitem [{\citenamefont {Page}(1993)}]{page1993}%
  \BibitemOpen
  \bibfield  {author} {\bibinfo {author} {\bibfnamefont {D.~N.}\ \bibnamefont {Page}},\ }\bibfield  {title} {\bibinfo {title} {{Average entropy of a subsystem}},\ }\href {https://doi.org/10.1103/PhysRevLett.71.1291} {\bibfield  {journal} {\bibinfo  {journal} {Phys. Rev. Lett.}\ }\textbf {\bibinfo {volume} {71}},\ \bibinfo {pages} {1291} (\bibinfo {year} {1993})}\BibitemShut {NoStop}%
\bibitem [{\citenamefont {Murciano}\ \emph {et~al.}(2022)\citenamefont {Murciano}, \citenamefont {Calabrese},\ and\ \citenamefont {Piroli}}]{murciano2022}%
  \BibitemOpen
  \bibfield  {author} {\bibinfo {author} {\bibfnamefont {S.}~\bibnamefont {Murciano}}, \bibinfo {author} {\bibfnamefont {P.}~\bibnamefont {Calabrese}},\ and\ \bibinfo {author} {\bibfnamefont {L.}~\bibnamefont {Piroli}},\ }\bibfield  {title} {\bibinfo {title} {{Symmetry-resolved Page curves}},\ }\href {https://doi.org/10.1103/PhysRevD.106.046015} {\bibfield  {journal} {\bibinfo  {journal} {Phys. Rev. D}\ }\textbf {\bibinfo {volume} {106}},\ \bibinfo {pages} {046015} (\bibinfo {year} {2022})}\BibitemShut {NoStop}%
\bibitem [{\citenamefont {Ghasemi}(2025)}]{ghasemi2025}%
  \BibitemOpen
  \bibfield  {author} {\bibinfo {author} {\bibfnamefont {M.}~\bibnamefont {Ghasemi}},\ }\bibfield  {title} {\bibinfo {title} {{Symmetry-resolved relative entropy of random states}},\ }\href {https://doi.org/10.1007/JHEP04(2025)090} {\bibfield  {journal} {\bibinfo  {journal} {Journal of High Energy Physics}\ }\textbf {\bibinfo {volume} {2025}},\ \bibinfo {pages} {1} (\bibinfo {year} {2025})}\BibitemShut {NoStop}%
\bibitem [{\citenamefont {Fukuhara}\ \emph {et~al.}(2013)\citenamefont {Fukuhara}, \citenamefont {Schau{\ss}}, \citenamefont {Endres}, \citenamefont {Hild}, \citenamefont {Cheneau}, \citenamefont {Bloch},\ and\ \citenamefont {Gross}}]{fukuhara2013}%
  \BibitemOpen
  \bibfield  {author} {\bibinfo {author} {\bibfnamefont {T.}~\bibnamefont {Fukuhara}}, \bibinfo {author} {\bibfnamefont {P.}~\bibnamefont {Schau{\ss}}}, \bibinfo {author} {\bibfnamefont {M.}~\bibnamefont {Endres}}, \bibinfo {author} {\bibfnamefont {S.}~\bibnamefont {Hild}}, \bibinfo {author} {\bibfnamefont {M.}~\bibnamefont {Cheneau}}, \bibinfo {author} {\bibfnamefont {I.}~\bibnamefont {Bloch}},\ and\ \bibinfo {author} {\bibfnamefont {C.}~\bibnamefont {Gross}},\ }\bibfield  {title} {\bibinfo {title} {{Microscopic observation of magnon bound states and their dynamics}},\ }\href {https://doi.org/10.1038/nature12541} {\bibfield  {journal} {\bibinfo  {journal} {Nature}\ }\textbf {\bibinfo {volume} {502}},\ \bibinfo {pages} {76} (\bibinfo {year} {2013})}\BibitemShut {NoStop}%
\bibitem [{\citenamefont {Christakis}\ \emph {et~al.}(2023)\citenamefont {Christakis}, \citenamefont {Rosenberg}, \citenamefont {Raj}, \citenamefont {Chi}, \citenamefont {Morningstar}, \citenamefont {Huse}, \citenamefont {Yan},\ and\ \citenamefont {Bakr}}]{christakis2023}%
  \BibitemOpen
  \bibfield  {author} {\bibinfo {author} {\bibfnamefont {L.}~\bibnamefont {Christakis}}, \bibinfo {author} {\bibfnamefont {J.~S.}\ \bibnamefont {Rosenberg}}, \bibinfo {author} {\bibfnamefont {R.}~\bibnamefont {Raj}}, \bibinfo {author} {\bibfnamefont {S.}~\bibnamefont {Chi}}, \bibinfo {author} {\bibfnamefont {A.}~\bibnamefont {Morningstar}}, \bibinfo {author} {\bibfnamefont {D.~A.}\ \bibnamefont {Huse}}, \bibinfo {author} {\bibfnamefont {Z.~Z.}\ \bibnamefont {Yan}},\ and\ \bibinfo {author} {\bibfnamefont {W.~S.}\ \bibnamefont {Bakr}},\ }\bibfield  {title} {\bibinfo {title} {{Probing site-resolved correlations in a spin system of ultracold molecules}},\ }\href {https://doi.org/10.1038/s41586-022-05558-4} {\bibfield  {journal} {\bibinfo  {journal} {Nature}\ }\textbf {\bibinfo {volume} {614}},\ \bibinfo {pages} {64} (\bibinfo {year} {2023})}\BibitemShut {NoStop}%
\bibitem [{\citenamefont {Friedenauer}\ \emph {et~al.}(2008)\citenamefont {Friedenauer}, \citenamefont {Schmitz}, \citenamefont {Glueckert}, \citenamefont {Porras},\ and\ \citenamefont {Sch{\"a}tz}}]{friedenauer2008}%
  \BibitemOpen
  \bibfield  {author} {\bibinfo {author} {\bibfnamefont {A.}~\bibnamefont {Friedenauer}}, \bibinfo {author} {\bibfnamefont {H.}~\bibnamefont {Schmitz}}, \bibinfo {author} {\bibfnamefont {J.~T.}\ \bibnamefont {Glueckert}}, \bibinfo {author} {\bibfnamefont {D.}~\bibnamefont {Porras}},\ and\ \bibinfo {author} {\bibfnamefont {T.}~\bibnamefont {Sch{\"a}tz}},\ }\bibfield  {title} {\bibinfo {title} {{Simulating a quantum magnet with trapped ions}},\ }\href {https://doi.org/10.1038/nphys1032} {\bibfield  {journal} {\bibinfo  {journal} {Nature Physics}\ }\textbf {\bibinfo {volume} {4}},\ \bibinfo {pages} {757} (\bibinfo {year} {2008})}\BibitemShut {NoStop}%
\bibitem [{\citenamefont {Neyenhuis}\ \emph {et~al.}(2017)\citenamefont {Neyenhuis}, \citenamefont {Zhang}, \citenamefont {Hess}, \citenamefont {Smith}, \citenamefont {Lee}, \citenamefont {Richerme}, \citenamefont {Gong}, \citenamefont {Gorshkov},\ and\ \citenamefont {Monroe}}]{neyenhuis2017}%
  \BibitemOpen
  \bibfield  {author} {\bibinfo {author} {\bibfnamefont {B.}~\bibnamefont {Neyenhuis}}, \bibinfo {author} {\bibfnamefont {J.}~\bibnamefont {Zhang}}, \bibinfo {author} {\bibfnamefont {P.~W.}\ \bibnamefont {Hess}}, \bibinfo {author} {\bibfnamefont {J.}~\bibnamefont {Smith}}, \bibinfo {author} {\bibfnamefont {A.~C.}\ \bibnamefont {Lee}}, \bibinfo {author} {\bibfnamefont {P.}~\bibnamefont {Richerme}}, \bibinfo {author} {\bibfnamefont {Z.-X.}\ \bibnamefont {Gong}}, \bibinfo {author} {\bibfnamefont {A.~V.}\ \bibnamefont {Gorshkov}},\ and\ \bibinfo {author} {\bibfnamefont {C.}~\bibnamefont {Monroe}},\ }\bibfield  {title} {\bibinfo {title} {{Observation of prethermalization in long-range interacting spin chains}},\ }\href {https://www.science.org/doi/10.1126/sciadv.1700672} {\bibfield  {journal} {\bibinfo  {journal} {Science advances}\ }\textbf {\bibinfo {volume} {3}},\ \bibinfo {pages} {e1700672} (\bibinfo {year} {2017})}\BibitemShut {NoStop}%
\bibitem [{\citenamefont {Ekert}\ \emph {et~al.}(2002)\citenamefont {Ekert}, \citenamefont {Alves}, \citenamefont {Oi}, \citenamefont {Horodecki}, \citenamefont {Horodecki},\ and\ \citenamefont {Kwek}}]{ekert2002}%
  \BibitemOpen
  \bibfield  {author} {\bibinfo {author} {\bibfnamefont {A.~K.}\ \bibnamefont {Ekert}}, \bibinfo {author} {\bibfnamefont {C.~M.}\ \bibnamefont {Alves}}, \bibinfo {author} {\bibfnamefont {D.~K.~L.}\ \bibnamefont {Oi}}, \bibinfo {author} {\bibfnamefont {M.}~\bibnamefont {Horodecki}}, \bibinfo {author} {\bibfnamefont {P.}~\bibnamefont {Horodecki}},\ and\ \bibinfo {author} {\bibfnamefont {L.~C.}\ \bibnamefont {Kwek}},\ }\bibfield  {title} {\bibinfo {title} {{Direct Estimations of Linear and Nonlinear Functionals of a Quantum State}},\ }\href {https://doi.org/10.1103/PhysRevLett.88.217901} {\bibfield  {journal} {\bibinfo  {journal} {Phys. Rev. Lett.}\ }\textbf {\bibinfo {volume} {88}},\ \bibinfo {pages} {217901} (\bibinfo {year} {2002})}\BibitemShut {NoStop}%
\bibitem [{\citenamefont {Islam}\ \emph {et~al.}(2015)\citenamefont {Islam}, \citenamefont {Ma}, \citenamefont {Preiss}, \citenamefont {Eric~Tai}, \citenamefont {Lukin}, \citenamefont {Rispoli},\ and\ \citenamefont {Greiner}}]{islam2015}%
  \BibitemOpen
  \bibfield  {author} {\bibinfo {author} {\bibfnamefont {R.}~\bibnamefont {Islam}}, \bibinfo {author} {\bibfnamefont {R.}~\bibnamefont {Ma}}, \bibinfo {author} {\bibfnamefont {P.~M.}\ \bibnamefont {Preiss}}, \bibinfo {author} {\bibfnamefont {M.}~\bibnamefont {Eric~Tai}}, \bibinfo {author} {\bibfnamefont {A.}~\bibnamefont {Lukin}}, \bibinfo {author} {\bibfnamefont {M.}~\bibnamefont {Rispoli}},\ and\ \bibinfo {author} {\bibfnamefont {M.}~\bibnamefont {Greiner}},\ }\bibfield  {title} {\bibinfo {title} {{Measuring entanglement entropy in a quantum many-body system}},\ }\href {https://doi.org/10.1038/nature15750} {\bibfield  {journal} {\bibinfo  {journal} {Nature}\ }\textbf {\bibinfo {volume} {528}},\ \bibinfo {pages} {77} (\bibinfo {year} {2015})}\BibitemShut {NoStop}%
\bibitem [{\citenamefont {Brydges}\ \emph {et~al.}(2019)\citenamefont {Brydges}, \citenamefont {Elben}, \citenamefont {Jurcevic}, \citenamefont {Vermersch}, \citenamefont {Maier}, \citenamefont {Lanyon}, \citenamefont {Zoller}, \citenamefont {Blatt},\ and\ \citenamefont {Roos}}]{brydges2019}%
  \BibitemOpen
  \bibfield  {author} {\bibinfo {author} {\bibfnamefont {T.}~\bibnamefont {Brydges}}, \bibinfo {author} {\bibfnamefont {A.}~\bibnamefont {Elben}}, \bibinfo {author} {\bibfnamefont {P.}~\bibnamefont {Jurcevic}}, \bibinfo {author} {\bibfnamefont {B.}~\bibnamefont {Vermersch}}, \bibinfo {author} {\bibfnamefont {C.}~\bibnamefont {Maier}}, \bibinfo {author} {\bibfnamefont {B.~P.}\ \bibnamefont {Lanyon}}, \bibinfo {author} {\bibfnamefont {P.}~\bibnamefont {Zoller}}, \bibinfo {author} {\bibfnamefont {R.}~\bibnamefont {Blatt}},\ and\ \bibinfo {author} {\bibfnamefont {C.~F.}\ \bibnamefont {Roos}},\ }\bibfield  {title} {\bibinfo {title} {{Probing R{\'e}nyi entanglement entropy via randomized measurements}},\ }\href {https://doi.org/10.18653/v1/2021.eacl-main.295} {\bibfield  {journal} {\bibinfo  {journal} {Science}\ }\textbf {\bibinfo {volume} {364}},\ \bibinfo {pages} {260} (\bibinfo {year} {2019})}\BibitemShut {NoStop}%
\bibitem [{\citenamefont {Castro-Alvaredo}\ and\ \citenamefont {Santamaría-Sanz}(2024)}]{castro2024}%
  \BibitemOpen
  \bibfield  {author} {\bibinfo {author} {\bibfnamefont {O.~A.}\ \bibnamefont {Castro-Alvaredo}}\ and\ \bibinfo {author} {\bibfnamefont {L.}~\bibnamefont {Santamaría-Sanz}},\ }\bibfield  {title} {\bibinfo {title} {{Symmetry-resolved measures in quantum field theory: A short review}},\ }\bibfield  {journal} {\bibinfo  {journal} {Modern Physics Letters B}\ }\textbf {\bibinfo {volume} {39}},\ \href {https://doi.org/10.1142/s0217984924300023} {10.1142/s0217984924300023} (\bibinfo {year} {2024})\BibitemShut {NoStop}%
\bibitem [{\citenamefont {Zyczkowski}\ and\ \citenamefont {Sommers}(2001)}]{zyczkowski2001}%
  \BibitemOpen
  \bibfield  {author} {\bibinfo {author} {\bibfnamefont {K.}~\bibnamefont {Zyczkowski}}\ and\ \bibinfo {author} {\bibfnamefont {H.-J.}\ \bibnamefont {Sommers}},\ }\bibfield  {title} {\bibinfo {title} {{Induced measures in the space of mixed quantum states}},\ }\href {https://iopscience.iop.org/article/10.1088/0305-4470/34/35/335} {\bibfield  {journal} {\bibinfo  {journal} {Journal of Physics A: Mathematical and General}\ }\textbf {\bibinfo {volume} {34}},\ \bibinfo {pages} {7111} (\bibinfo {year} {2001})}\BibitemShut {NoStop}%
\end{thebibliography}

\end{document}